\documentclass[fleqn,usenatbib]{mnras}

\usepackage[T1]{fontenc}

\DeclareRobustCommand{\VAN}[3]{#2}
\let\VANthebibliography\thebibliography
\def\thebibliography{\DeclareRobustCommand{\VAN}[3]{##3}\VANthebibliography}

\usepackage{graphicx}	
\usepackage{amsmath}	
\usepackage{amssymb}	
\usepackage{pdflscape}
\usepackage{tablefootnote}

\usepackage{newtxtext,newtxmath}

\title[All-purpose photo-$z$s for LS DR8]{All-purpose, all-sky photometric redshifts for the Legacy Imaging Surveys Data Release 8}

\author[K. J. Duncan]{
Kenneth J. Duncan$^{1}$\thanks{E-mail: kdun@roe.ac.uk}
\\
$^{1}$ Institute for Astronomy, University of Edinburgh, Royal Observatory, Blackford Hill, Edinburgh, EH9 3HJ, UK
}

\date{Accepted 2022 March 3. Received 2022 February 22; in original form 2021 November 23}

\pubyear{2022}

\begin{document}
\label{firstpage}
\pagerange{\pageref{firstpage}--\pageref{lastpage}}
\maketitle

\begin{abstract}
In this paper we present photometric redshift (photo-$z$) estimates for the Dark Energy Spectroscopic Instrument (DESI) Legacy Imaging Surveys, currently the most sensitive optical survey covering the majority of the extra-galactic sky. 
Our photo-$z$ methodology is based on a machine-learning approach, using sparse Gaussian processes augmented with Gaussian mixture models (GMMs) that allow regions of parameter space to be identified and trained separately in  a purely data-driven way.
The same GMMs are also used to calculate cost-sensitive learning weights that mitigate biases in the spectroscopic training sample.
By design, this approach aims to produce reliable and unbiased predictions for all parts of the parameter space present in wide area surveys.
Compared to previous literature estimates using the same underlying photometry, our photo-$z$s are significantly less biased and more accurate at $z > 1$, with negligible loss in precision or reliability for resolved galaxies at $z < 1$.
Our photo-$z$ estimates offer accurate predictions for rare high-value populations within the parent sample, including optically selected quasars at the highest redshifts ($z > 6$), as well as X-ray or radio continuum selected populations across a broad range of flux (densities) and redshift.
Deriving photo-$z$ estimates for the full Legacy Imaging Surveys Data Release 8, the catalogues provided in this work offer photo-$z$ estimates predicted to be high quality for $\gtrsim9\times10^{8}$ galaxies over $\sim 19\,400\,\text{deg}^{2}$ and spanning $0 < z \lesssim 7$, offering one of the most extensive samples of redshift estimates ever produced.
\end{abstract}




\section{Introduction}\label{sec:intro}
Photometric redshift estimates (photo-$z$s) represent a fundamental ingredient in studies of both galaxy and active galactic nuclei (AGN) evolution, and in an increasing number of cosmological experiments.
Broadly, methods of estimating photo-$z$s fall into two categories: template fitting approaches that fit model or empirical template libraries to the observed photometry \cite[most effective and popular for deep survey fields with large numbers of available filters;][]{Dahlen:2013eu}, and machine learning (ML) approaches that use classification or regression algorithms trained on reference samples to predict new redshifts for the science sample of interest \citep[e.g.][]{Collister:2004fx, CarrascoKind:2013kd, CarrascoKind:2014gb}.
Thanks to the ever increasing depth and fidelity of photometric surveys, the samples of optical sources for which these photo-$z$ estimates are required has grown significantly - requiring an increasing dependence on machine learning photo-$z$ approaches due to their potential efficiency and scalability \citep[see e.g.][]{2019NatAs...3..212S}.
In addition to the disadvantage in speed, reliable template-fitting photo-$z$s are more reliant on robust photometry and good photometric calibration \citep{Hildebrandt:2012du}, as well the use of appropriate template libraries that are able to encompass the range of colours present in the population of interest \citep[e.g.][]{2019MNRAS.489.3351B, 2017ApJ...850...66A} - something which becomes increasingly challenging as sample sizes grow.

Conversely, ML methods are less impacted by systematic errors within photometry that can limit template fitting as they do not rely on physically consistent models.
Rather, ML methods learn the correlations between the apparent colour or magnitude and redshift for the specific dataset in question. 
Furthermore, they are also able to incorporate additional valuable information such as source size and morphology \citep{2018MNRAS.475..331G, 2020ApJ...888...83W} that can break redshift degeneracies. 
However, given the requirement for training samples, machine learning photo-$z$ estimates perform best for populations where large spectroscopic samples are available for training and testing. 
Many machine learning photo-$z$ estimates are therefore tailored towards the particular population of interest by design \citep{2007MNRAS.375...68C,2012ApJ...749...41B,2018MNRAS.478..592H}.
Even in the case where a broad range of populations are included within the training samples, the optimisation process for many of the regression or classification algorithms employed will be strongly weighted towards the most densely populated regions of the training space. 
A natural consequence of this is that estimates for rare or extreme populations can be extremely biased or imprecise \citep[see e.g.][]{2019PASP..131j8004N,2020A&A...644A..31E}.  
Alternatively, studies of samples which span a broad range of optical properties must rely on multiple different sets of photo-$z$ estimates for different sub-populations, each with their own systematic biases and limitations.

These potential limitations are particularly relevant for samples selected through other wavelength regimes, such as radio continuum and X-ray surveys, where the range of optical properties can be extremely broad.
For example, radio-selected samples from the new generation of wide area radio continuum surveys with MeerKAT \citep{Booth:2009wx}, the Australian SKA Pathfinder \citep[ASKAP;][]{2007PASA...24..174J} and the Low Frequency Array \citep[LOFAR;][]{vanHaarlem:2013gi} can range from massive elliptical galaxies at low redshift \citep{2019A&A...622A..17S} through to luminous quasars at $z > 5$ \citep{2021arXiv211006222G}.
\citet[][see also \citeauthor{Duncan:21}~\citeyear{Duncan:21}]{2019A&A...622A...3D} tried to solve this problem using a `hybrid' machine learning and template fitting approach over 400 deg$^{2}$ - with the template estimates contributing to good estimates for a wide range of source properties, while machine learning estimates provided better precision and reduced bias for the population where larger training samples were available \citep[see][for a similar approach to combining ML and template estimates]{2017MNRAS.466.2039C}.
While the hybrid approach provides photo-$z$ estimates with sufficient quality to enable a broad range of science \citep[e.g.][]{2019MNRAS.488.2701M,2019A&A...622A..12H, Smith:2020di}, the processing requirements of the template fitting estimates prohibit extension of this method to much larger datasets.

Extensions of data driven photo-$z$ methods to better cope with a diverse range of populations have been proposed.
\citet{2020MNRAS.498.5498H} present one solution, whereby unsupervised clustering of sources is used to divide the population into distinct regions of parameter space (specifically, colour-magnitude space), with machine learning photo-$z$ estimates then trained separately for each region.
By treating each region of parameter space independently, the resulting photo-$z$ estimates are significantly less biased and more robust than comparable estimates that include the full population within a single estimator.
However, while promising, the methods outlined in \citet{2020MNRAS.498.5498H} have so far only been applied to relatively small, but deep photometric datasets.

In this paper, we aim to build upon the methods of \citet{2019A&A...622A...3D}, \citet{Duncan:21} and \citet{2020MNRAS.498.5498H} to provide photo-$z$ estimates for the deepest photometric survey available over the full extragalactic sky, namely the Dark Energy Spectroscopic Instrument (DESI) Legacy Imaging Surveys \citep[][colloquially known as the `Legacy Surveys']{2019AJ....157..168D}.
While photo-$z$ estimates have been produced for the Legacy Surveys eighth data release (LS DR8 hereafter) and previous data releases, these estimates have crucially been limited to subsets of the optical population.
For example, \citet{2019ApJS..242....8Z} presents photo-$z$ and stellar mass estimates for Legacy Surveys DR6 and 7 (since updated to also include DR8).
By construction these estimates are limited to the redshift range $z < 1$, with morphological cuts imposed to exclude unresolved sources (i.e. those modelled by a point-spread function) from the analysis.
Similarly, \citet{2021MNRAS.501.3309Z} presents photo-$z$ estimates primarily targeted at the luminous red galaxy (LRG) population, with strict magnitude and colour cuts imposed to the limit the analysis to the target population.
In both cases, the selection criteria are well motivated by the data and respective scientific goals and the resulting photo-$z$ estimates are of a high quality for the chosen sub-populations.
However, given the explicit colour, morphology and redshift ranges for which the \citet{2019ApJS..242....8Z} and \citet{2021MNRAS.501.3309Z} photo-$z$s are designed, there remain key subsets of the optical population without valid photo-$z$ estimates, most notably unresolved optical sources at all redshifts (e.g. quasars) as well as resolved optical sources with significant AGN contributions and wider $z > 1$ populations.

The goal of this paper is therefore to provide a set of photo-$z$ estimates that encompass all populations for which robust photo-$z$s can be derived (given the observational limits of the photometry), regardless of redshift, colour or morphology.
Our photo-$z$ methodology will build upon the approach presented in \citet{2020MNRAS.498.5498H}, using the data itself to divide the empirical photo-$z$ training and prediction into different regions of optical parameter space with the goal of both improved results and computational performance. 

The remaining sections of this paper are set out as follows.
In Section~\ref{sec:data}, we present details of the optical catalogues used in this analysis, the processing and homogenisation steps applied to them and the spectroscopic redshift samples used for photo-$z$ training and testing.
Section~\ref{sec:method} then outlines the photo-$z$ methodology employed, including testing and analysis of the variety of possible approaches explored and details of the final photo-$z$ training and estimation steps. 
Next, Section~\ref{sec:results} presents a quantitative analysis of the resulting photo-$z$ estimates, including their performance as a function of optical properties. 
We also analyse the photo-$z$ performance relative to existing estimates using the Legacy DR8 catalogues and for key subsets of the optical population (radio continuum and X-ray).
Finally, Section~\ref{sec:catalogues} provides details of the photo-$z$ catalogues presented in this analysis before Section~\ref{sec:summary} presents a summary of our work.
Throughout this paper, all magnitudes are quoted in the AB system \citep{1983ApJ...266..713O} unless otherwise stated. We also assume a $\Lambda$ Cold Dark Matter cosmology with $H_{0} = 70$ km\,s$^{-1}$\,Mpc$^{-1}$, $\Omega_{m}=0.3$ and $\Omega_{\Lambda}=0.7$.

\section{Observational Data}\label{sec:data}
\subsection{Photometric data}

\subsubsection{DESI Legacy Imaging Surveys}\label{sec:ls_data}
Initially motivated by target selection requirements for the forthcoming DESI spectroscopic survey, the Legacy Surveys comprise a set of optical photometric surveys providing imaging and source catalogues in the $g$, $r$ and $z$ bands.
Specifically, the Legacy Surveys consists of the Beijing Arizona Sky Survey \citep[BASS, $g/r$;][]{2017AJ....153..276Z} and the Mayall $z$-band Legacy Survey \citep[MzLS, $z$;][]{2016AAS...22831702S} providing imaging at declinations $\delta \gtrsim 32$ in the North Galactic cap (LS DR8 North). 
Imaging over the remainder of the extragalactic sky (LS DR8 South) is provided by the Dark Energy Camera Legacy Survey (DeCaLS, PIs: D. Schlegel and A. Dey), supplemented by all publicly available imaging in the $g$, $r$ and $z$ bands \citep[predominantly the Dark Energy Survey,][]{2005astro.ph.10346T}.
Combined, the optical imaging in LS DR8 covers over 19\,000 deg$^{2}$ and is typically at least $\sim1$ mag deeper than the Panoramic Survey Telescope and Rapid Response System (Pan-STARRS) 3$\pi$ Steradian Survey \citep{2016arXiv161205560C} in each band, reaching median $5\sigma$ depths  $m_{g}\sim 24.3-24.9$, $m_{r}\sim 23.7-24.3$, and $m_{z}\sim 23.3-23.4$.

In addition to photometry from the dedicated optical imaging, the Legacy Surveys catalogues also incorporate mid-infrared (mid-IR) photometry from the Wide-field Infrared Survey Explorer mission \citep[WISE;][]{Wright:2010in} at 3.4, 4.6, 12 and 22\,$\mu$m (W1, W2, W3 and W4 respectively).
WISE photometry is provided for all optically detected sources based on the combined data from the cryogenic and post-cryogenic \citep[NEOWISE;][]{2011ApJ...731...53M} and NEOWISE-reactivation \citep[or NEOWISER;][]{2017AJ....153...38M, 2017AJ....154..161M} missions.

Full details of photometric calibration, source detection and photometry are provided in \citet[][and references therein]{2019AJ....157..168D}, but here we summarise the key details.
After self-consistent astrometric and photometric calibration, source detection is performed using the three individual bands and two combined $grz$ stacks (an un-weighted stack optimised for "flat" spectral energy distributions, and a weighted sum optimised for red sources).
In all five detection images, sources are selected using a simple thresholding algorithm, retaining unique detections above $6\sigma$ in any individual band.
Photometry is then performed using \textit{The Tractor} \citep{2016ascl.soft04008L}, which simultaneously models each source across all filters using a set of parametric light-profiles. 
Specifically, \textit{The Tractor} models sources as either point-sources (i.e. with a delta function; $\texttt{type} =$ `PSF'), or as resolved sources with either a round exponential profile (`REX'), an exponential disk (`EXP'), a de Vaucouleurs $r^{-1/4}$ power-law (`DEV'), or a "composite" model with both an exponential and de Vaucouleurs component (`COMP').
During fitting of $grz$ photometry, the respective models are convolved with the corresponding point-spread function (PSF), with the PSF for individual exposures computed for each CCD separately using stars in the field \citep[see][for further details]{2019AJ....157..168D}.
\textit{The Tractor} assumes that the same model applies across all optical bands, i.e. if a source is spatially extended, the same light-profile is fitted to all images so that shape and size measurements are consistent.
Therefore, in addition to providing robust total flux measurements, morphological type, shape and size information is available for every source in the LS DR8 catalogue.

Mid-IR photometry is performed in a similar way, with \textit{The Tractor} forcing the location and shape of sources based on the optical information and convolving with the respective WISE PSF to model the flux of the sources in each WISE band.
Using the $grz$ information allows for deblending of confused WISE sources, reliably probing to significantly fainter mid-IR fluxes than possible with blind mid-IR selected photometric catalogues \citep{Wright:2010in}.
The additional mid-IR sensitivity provided by this deblending is particularly valuable for photo-$z$ estimation with LS DR8 due to the small number of optical filters available.

All of the Legacy surveys source detection and photometry is performed in small $0.25^{\circ} \times  0.25^{\circ}$ regions known as ``bricks".
For this work, we restrict our analysis to only those bricks with at least one exposure in all three optical bands.
When both downloading the LS DR8 catalogue data and using them for photo-$z$ prediction, we make the decision to group bricks into larger regions using  Hierarchical Equal Area isoLatitude Pixelation \citep[\textsc{HEALPix};][]{2005ApJ...622..759G}\footnote{\href{http://healpix.sourceforge.net}{http://healpix.sourceforge.net}}.
Bricks are grouped based on the \textsc{HEALPix} order 3 pixel (with area $\approx 53.7$ deg) in which the right ascension and declination of the brick centre lies.
This results in 123 and 349 distinct \textsc{HEALPix} regions for LS DR8 North and South respectively, which together cover the total footprint of $\sim19\,400$ deg$^{2}$ (note that there is overlap between North and South and not all \textsc{HEALPix} regions are 100\% filled, particularly around the survey edges).
After restricting to primary sources for which photometric measurements are available, the final source catalogues for analysis comprise 323\,213\,867 sources in LS DR8 North and 1\,252\,523\,992 sources in the South.

\subsubsection{$\emph{asinh}$ magnitudes}
The LS DR8 photometric catalogues provide optical ($grz$) and WISE total flux density measurements (and associated uncertainties) for all sources.
Previous studies, however, have shown that photo-$z$ estimates derived using magnitudes yield improved results for methodologies such as those employed in this study \citep{2020MNRAS.498.5498H,Duncan:21}.
To enable training and estimation in log-flux space whilst also allowing for the possibility of non-detections and negative flux measurements,  we follow the approach outlined in \citet{Duncan:21} and derive \emph{asinh} magnitudes \citep[][also colloquially known as `luptitudes']{1999AJ....118.1406L} for the purposes of photo-$z$ training and estimation.
For a given flux, $f$ (with a flux zeropoint $f_{0} = 3631~\rm{Jy}$), our \emph{asinh} magnitudes are defined as
\begin{equation}
	 m = \frac{-2.5}{\log(10)} \times \sinh^{-1}\left ( \frac{f/f_{0}}{2b} \right ) + \log(b)
\end{equation}
where the softening parameter, $b$, is unit-less (normalised by $1/f_{0}$). As defined in \citet{1999AJ....118.1406L}, the optimal choice for $b$ is approximately equal to the noise in the flux.
The corresponding magnitude uncertainties are then given by 
\begin{equation}
    \sigma_{m} =  \frac{-2.5}{\log(10)} \times \frac{(\sigma_{f}/|f|)}{ \sqrt{\left(1 + (2b / (f/f_{0}))^{2}\right )}}
\end{equation}
where $\sigma_{f}$ is the 1-$\sigma$ flux uncertainty derived during model-fitting photometry (converted from inverse variance; \texttt{flux\_ivar\_[g/r/z]}).
When calculating \emph{asinh} magnitudes, we correct both the flux and flux uncertainties for Galactic extinction based on the transmission provided in the LS DR8 catalogues \citep[which combine the measured $E(B-V)$ of][ and the respective extinction coefficients, $A_{\lambda}/E(B-V)$, for each filter]{1998ApJ...500..525S}.

Previously \citet{2019MNRAS.489..820B} have shown that photo-$z$ estimates derived using \emph{asinh} magnitudes are not sensitive to the exact definition of softening parameter ($b$) assumed, provided the data are of similar depth across the field in question.
However, due to the variation in depth between LS DR8 North and South optical catalogues and the significant variation within each dataset, the assumption of homogeneity and a single value for $b$ for each filter is not justified.
We therefore make the decision to define the softening parameter $b$ based on the local 1$\sigma$ PSF depth for each source in the respective bands (e.g. $b_{g} = \sigma_{\text{PSF},g}/f_{0}$, where $\sigma_{\text{PSF},g}$ is \texttt{psfdepth\_g} from the LS DR8 catalogue converted to Jy).
Based on direct comparison of photo-$z$ estimates derived for sources in common between both LS DR8 North and South (which differ in optical depth, and hence softening parameter), we find the estimates are not sensitive to the exact choice of $b$ (Appendix~\ref{app:pz_offsets}), in line with previous studies.

\subsubsection{Data homogenisation}\label{sec:data_homogenisation}
 Although both Legacy North and South imaging and catalogues have been processed and calibrated following the same procedures, the difference in effective filter response curves between the sets of $g/r/z$ observations means that the two datasets are not completely homogenous. 
To avoid systematic biases or offsets between datasets (and their associated regions of the sky), photo-$z$ training and estimation must either be done separately for the two catalogues \citep[as in][]{2021MNRAS.501.3309Z}, or colour corrections applied to homogenise to a consistent photometric system.
While spectroscopic training samples sufficient for good photo-$z$ estimates are available within both datasets, Legacy South benefits from significantly improved training samples for complete magnitude selected samples (due to the presence of equatorial deep fields).

\begin{figure*}
\centering
 \includegraphics[width=0.95\textwidth]{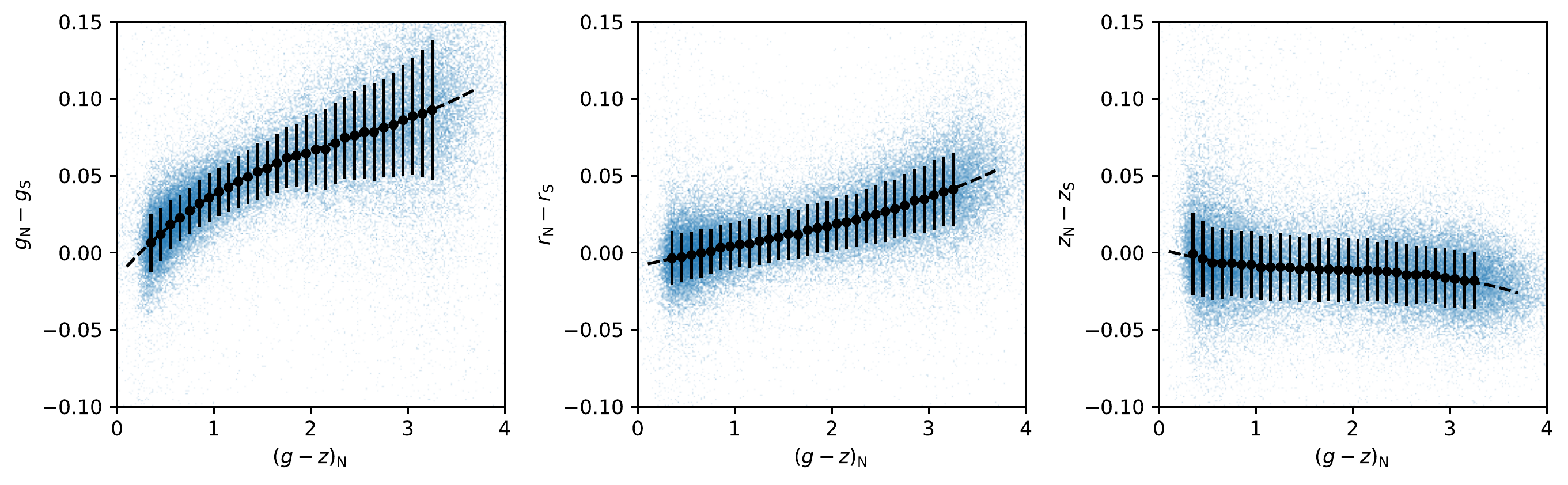}
 \caption{Observed magnitude offsets between Legacy North and South datasets as a function of $g - z$ colour for a sample of $\sim9\times 10^{5}$ objects where both datasets are available. Blue dots demonstrate individual measurements used in the analysis. Black circles show the 3$\sigma$-clipped mean offset in bins of $(g-z)_{\text{N}}$, with error bars showing the corresponding sigma-clipped standard deviation. For each filter ($g / r/ z$), the trend is fitted with a cubic polynomial (dashed line).}
 \label{fig:north-south}
\end{figure*}

Therefore, in this work, we take the approach of homogenising the photometric datasets and using the full spectroscopic training sample to train photo-$z$ models that can be applied uniformly to both hemispheres.
Our colour corrections exploit the fact that there is a significant region of overlap between the two sets of imaging and catalogues.
After selecting a subset of \textit{The Tractor} bricks from the overlapping region, we positionally cross-matched the two sets of catalogues, keeping only those sources with matches within 0.35\arcsec (the pixel scale).
For a large set of bright, high S/N point-sources ($>10\sigma$ detections in each dataset), we then measured the \emph{asinh} magnitude offset between the measurements through each pair of filters as a function of observed $g - z$ colour in the Legacy North catalogue.
The resulting magnitude offsets for each of the three optical bands as a function of colour are illustrated in Fig.~\ref{fig:north-south}, both for individual sources (blue points) and the robust 3-$\sigma$ clipped mean and standard deviation within bins of $(g - z)_{\text{N}}$.  

The largest magnitude offsets between North and South catalogues are observed in their respective $g$-band filters, with the systematic offset reaching $+0.1$ dex for the reddest sources.
Smaller offsets are observed in the $r$ and $z$ bands, reaching $\approx0.03$ and $\approx -0.02$ dex offsets respectively.
The more significant offsets observed for the $g$ filters are line with expectations based on the differences in the respective filter response curves \citep[see Fig.~3 of][]{2019AJ....157..168D}.

To homogenise the photometry before photo-$z$ training and estimation, we parametrise the trend as a function of $(g - z)_{\text{N}}$ using simple cubic polynomial fits (black dashed line in Fig.~\ref{fig:north-south}).
All sources in the Legacy North photometric catalogues are then corrected to match the Legacy South system based on their observed $(g - z)_{\text{N}}$ colour, with the observed colours clipped to the range $0.3 < (g - z)_{\text{N}} < 3.3$ to prevent overcorrection due to extreme (potentially unphysical) colours resulting from photometric scatter in lower S/N sources.
Given the average measured colour within the full catalogued population is $ (g - z)_{\text{N}} = 1.59$, the average magnitude offsets applied to the Legacy North photometry ar 0.056, 0.012 and -0.010 in the $g / r/ z$ bands respectively. 
The impact (or lack thereof) of any residual colour systematics on the final resulting photo-$z$ estimates are explored in further detail in Appendix~\ref{app:pz_offsets}.
 
\subsection{Spectroscopic training data}\label{sec:data-specz}
Thanks to the large spatial coverage of the combined Legacy photometric datasets, extensive spectroscopic training samples are available over a very broad range of parameter space. 
We compile our full training sample from two primary datasets.
Firstly, we make use of the combined samples presented in the 14th data release of the Sloan Digital Sky Survey \citep[SDSS DR14;][]{2018ApJS..235...42A}, which includes the SDSS main galaxy sample \citep{2002AJ....124.1810S} as well as the subsequent Baryon Oscillation Spectroscopic Survey \citep[BOSS][]{2013AJ....145...10D} and the Extended BOSS \citep{2016AJ....151...44D} samples. 
For any SDSS spectroscopic source which was also in the separate DR14 quasar catalogue \citep[][DR14Q hereafter]{2018A&A...613A..51P}, we use the redshift measurement provided by DR14Q in place of the standard SDSS pipeline redshift.
For both samples, the SDSS catalogues are joined with the LS DR8 photometry using the row matched catalogues provided by Legacy Surveys, which provides the closest matching sources up to a maximum separation of 1.5\arcsec.
Since the combined SDSS spectroscopic sample extends to over 2 million individual sources, for our training and testing we make use of a random subset of 50\% of the SDSS sample, reserving the second half for potential future testing of the spatial variation in photo-$z$ quality or bias.

We next supplement the large samples provided by SDSS with an additional sample of spec-$z$s compiled over a large number of deeper survey fields for the Herschel Extragalactic Legacy Project \citep[HELP][]{2021arXiv210505659S}.
The full HELP spectroscopic redshift sample consists of publicly available spectroscopic redshifts from 101 individual spectroscopic surveys, with all input catalogues homogenised to consistent formats and coordinate systems.\footnote{The spec-$z$ compilation can be downloaded in its entirety or queried through standard Virtual Observatory tools at the \href{https://herschel-vos.phys.sussex.ac.uk/browse/specz/q}{HELP Virtual Observatory Server}.}
As part of the HELP compilation procedure, whenever multiple spec-$z$ measurements exist for the same individual source, the `best' available measurement has been retained based on visual inspection and duplicate entries have been removed \citep[see][for further details]{2021arXiv210505659S}. 

The HELP spec-$z$ sample was cross-matched to LS DR8 using simple nearest neighbour match, employing a more conservative maximum separation than employed for the SDSS samples ($<1\arcsec$).
In total, 618\,606 sources in the HELP compilation are matched with the LS DR8 photometry catalogues.
However, this number includes a fraction of sources with unreliable spectroscopic redshifts, as well as some duplication of sources already contained in the SDSS sample.
For any source which is present in both the SDSS and HELP samples, we therefore remove the duplicate source to prevent over-weighting the source in training, retaining the spec-$z$ measurement provided by HELP.
Additionally, we include only sources deemed to have reliable redshift measurements, with HELP sources requiring a quality flag of 3 or higher, corresponding to ``good'', ``reliable'' or high confidence (>90\% confidence) redshifts depending on individual surveys definition \citep[see Section. 4.5 of][for additional details]{2021arXiv210505659S}. 
For SDSS spectroscopic redshifts, we also require all training sources to have no redshift warning flags (\texttt{ZWARNING} = 0).

Finally, to ensure that our training sample is robust and as complete as possible for quasars out to the highest redshifts, we supplement the SDSS quasar samples with the compilation of $z > 5$ quasars presented by \cite{2020MNRAS.494..789R}.
The Very High-$z$ Quasar (VH$z$Q) catalogue\footnote{\url{www.github.com/d80b2t/VHzQ}} consists of 488 $z>5$ quasars with robust spectroscopic confirmations compiled from the literature, ensuring accurate positions with consistent astrometry.
Matching to the LS DR8 catalogues with the same 1\arcsec\, limit used for the HELP sample, we find matches for 375 sources extending out to redshifts as high as $z = 7.0842$.
    
  To ensure a clean training dataset we apply additional photometric quality criteria based on the Legacy photometry catalogues.
We require $\texttt{fracflux\_[g/r/z]}$ (the profile-weighted fraction of the measured flux from other sources divided by the total flux) in each optical bands to be $ < 0.2$, ensuring that the source is not significantly impacted by blending with neighbouring sources. 
  Similarly, $\texttt{fracmasked\_[g/r/z]}$ (the profile-weighted fraction of pixels masked from all observations of this object) is required to be $ < 0.1$, such that the source is not affected by imaging artefacts.
  Additionally, we require $\texttt{maskbits} = 0 $, such that the source does not overlap with regions included in the Legacy bitmasks.\footnote{See \href{https://www.legacysurvey.org/dr8/bitmasks/}{https://www.legacysurvey.org/dr8/bitmasks/} for additional information on LS DR8 bitmask definitions.}
Together, these criteria ensure that the training sample is free from significant source blending or contamination from bright nearby sources (e.g. artefacts around bright stars or extended galaxies) that would impact the observed colours, and hence bias the resulting photo-$z$ estimates.

\begin{figure}
\centering
 \includegraphics[width=1.0\columnwidth]{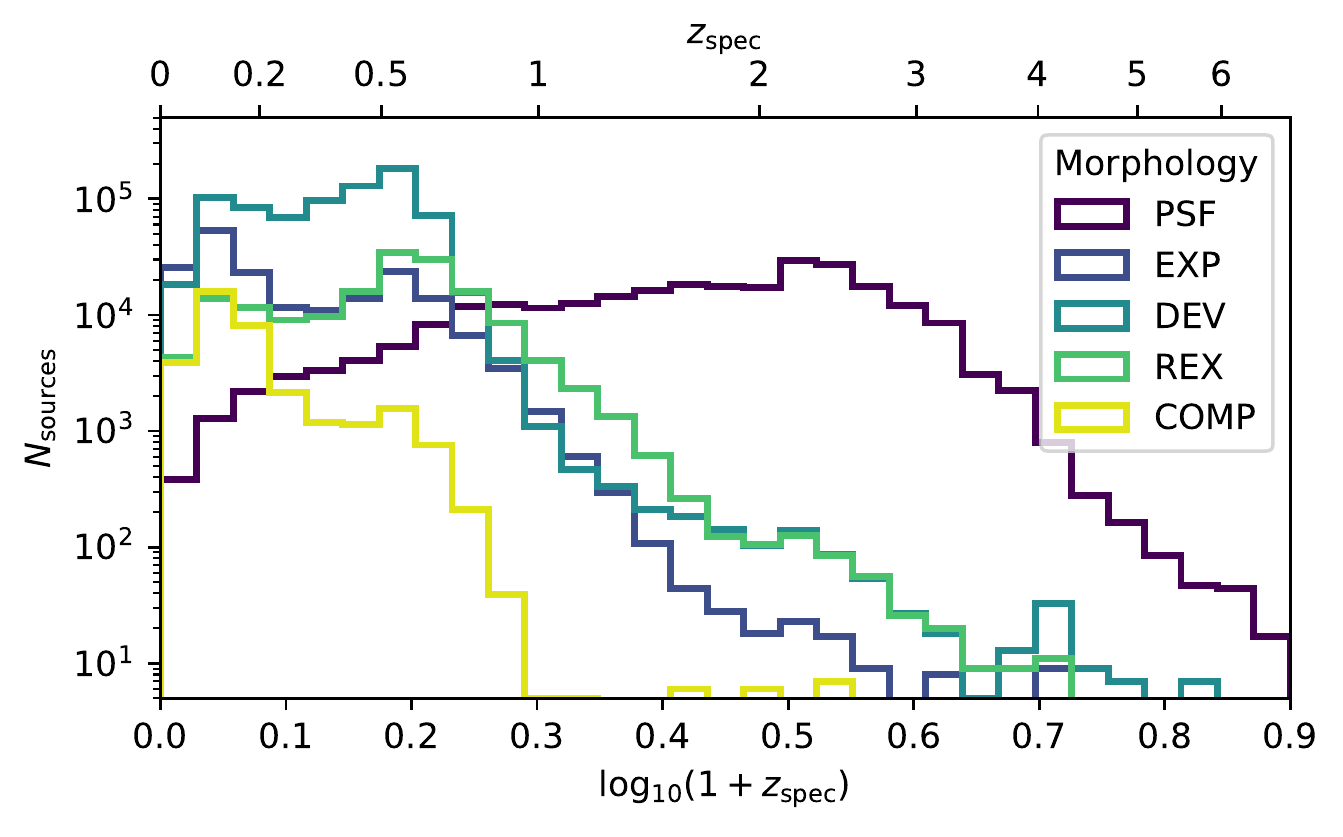}
 \caption{Redshift distribution of the final spectroscopic training sample, split by optical morphology classification within the LS DR8 catalogue (\texttt{type}). As would be expected given the optical photometry resolution and depths, resolved sources (EXP, DEV, REX, COMP) are largely confined to $z < 1$, while unresolved sources (PSF; predominantly quasars) extend to $z > 6$.}
 \label{fig:specz_training_dist}
\end{figure}

Since automated source classification and redshifting pipelines employed for large surveys can never be 100\% reliable, small numbers of stars can contaminate even notionally `robust' spec-$z$s samples.
As a final cut, we exclude from the spectroscopic training sample potential stellar contaminants that exhibit either significant measured proper motions \citep[Gaia parallax S/N > 3;][]{2016A&A...595A...1G} or have colours consistent with stars ($P_{\text{star}} > 0.2$; see Section~\ref{sec:star-gal}).
In Fig.~\ref{fig:specz_training_dist} we present the redshift distribution for the final spectroscopic sample, split by best-fitting morphological type from the LS DR8 photometry. 
The total sample statistics are also summarised in Table~\ref{tab:specz_summary}.

\begin{table}
\centering
\caption{Total sample sizes for the spectroscopic training, validation and test subsets used in the photo-$z$ estimation. Training and validation samples are defined at the point of training (see Section~\ref{sec:method}), with a ratio of 3:1.}
\begin{tabular}{lrr}
\hline
Morphology & Training \& Validation & Test \\
\hline
PSF & 260\,930 & 65\,237 \\
EXP & 188\,792 & 47\,202 \\
DEV & 775\,309 & 193\,832 \\
REX & 162\,426 & 40\,612 \\
COMP & 35\,032 & 8\,762 \\
\hline
All & 1\,422\,489 & 355\,645 \\
\hline
\end{tabular}\label{tab:specz_summary}
\end{table}

\subsection{Star-QSO classification}\label{sec:star-gal}
Although robust classification of photometric point-sources in the LS DR8 catalogues is not a primary goal of this analysis, the inclusion of known stars within various steps in our methodology could negatively impact both the accuracy and precision of estimates for the extra-galactic point-source population.
We therefore first use the existing spectroscopic samples to train a star-galaxy classifier using the colour information available in the optical catalogues.
\begin{figure*}
\centering
 \includegraphics[width=0.92\textwidth]{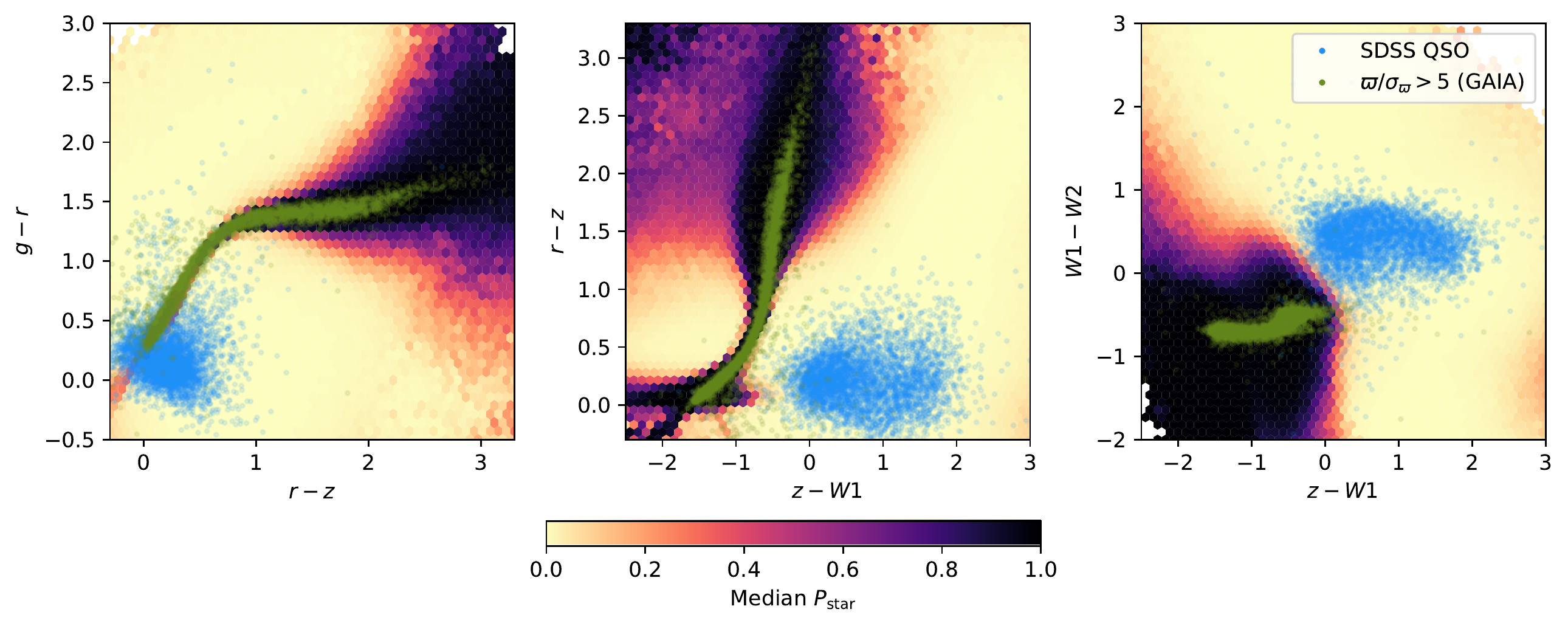}
 \caption{Median star probability ($P_{\text{star}}$, colour-scale) as a function of optical and mid-infrared colour in a sample of $\sim3.5\times10^{6}$ sources based on the GMM classification. A minimum of 5 sources per cell are required for plotting (with empty cells shown in white). To illustrate the colours of key samples of interest, also plotted are a random subset of 5000 known SDSS QSOs (blue points) and 5000 Legacy catalogue sources identified as stars independently from their colour based on high-significance parallax measurements from \textit{Gaia} (green points). The importance of WISE observations in cleanly separating the stellar population from QSOs can be seen in the middle and right panels.}
 \label{fig:star_prob}
\end{figure*}

Motivated by previous successful approaches \citep{2011ApJ...729..141B,2012ApJ...749...41B}, we employ a Gaussian Mixture Model (GMM) to derive the densities of different source classes within the available photometric colour-space.
GMMs assume that a given dataset can be modelled by a mixture of a finite number of multi-dimensional Gaussian distributions and can be used both for the purposes of clustering to identify sub-populations within the sample, or as in this case to flexibly modelling the probability density of a distribution of data (i.e. the colours of stars).
To derive the star-QSO classification we construct GMMs for two different source classes. 
Firstly, we use morphological point-sources (\texttt{type} = `PSF') from the spec-$z$ training sample to model the colour-space of the galaxy/QSO population. 
Note that this sample includes all PSF sources, regardless of whether they were explicitly targeted as QSOs.
Next, we then use sources from the SDSS DR14 spectroscopic sample \citep{2018ApJS..235...42A} classified as `STAR' to model the stellar population.
We model each GMM using four colours derived from the LS DR8 catalogues; $g - r$, $r - z$, $z - W1$, $W1 - W2$, with 20 components (see Section~\ref{sec:method-divide} for further discussion of the choice of how many components are used to construct the GMM).
The resulting star probability is then defined as
\begin{equation}
    P_{\text{star}}(\mathbf{\hat{m}}) \propto \frac{\mathcal{L}_{\text{star}}(\mathbf{\hat{m}})}{\mathcal{L}_{\text{star}}(\mathbf{\hat{m}}) + \mathcal{L}_{\text{QSO}}(\mathbf{\hat{m}})}
\end{equation}\label{eq:pstar}
\noindent where $\mathcal{L}(\mathbf{\hat{m}})$ is the probability density in a given GMM for a set of observed colours, $\mathbf{\hat{m}}$.
Note that strictly, to derive a true probability Eq.~\ref{eq:pstar} should be amended to incorporate prior probabilities on the relative source densities of the star and quasar populations \citep[also including the dependence on observed magnitude;][]{2011ApJ...729..141B}.
However, for the purposes of this work, it is sufficient to identify and remove the likely stellar population as the photo-$z$ estimates themselves can be used to further identify true QSO.
An additional caveat to our star-QSO separation procedure is that it is naturally biased towards stellar types present within the SDSS DR14 training sample and therefore may be incomplete for some key populations with colours similar to some QSO populations (e.g. cool dwarf stars and high redshift QSOs). 
The full stellar training sample does however include stars initially targeted as quasar candidates based on their optical colours \citep{2015ApJS..219...12A}, providing valuable training data in the key regions of optical parameter space where the two populations are most similar (and where infrared colours can offer key information).
 
To illustrate the effectiveness of our star-QSO classifications for bright point sources, in Fig.~\ref{fig:star_prob} we show how the derived $P_{\text{star}}$ varies as a function of source colour.
We first calculate $P_{\text{star}}$ for a sample of $\sim3.5\times10^{6}$ PSF sources selected randomly from the LS DR8 catalogues.
For three projections of the 4-dimensional colour space used in the GMM, we then calculate the average $P_{\text{star}}$ as a function of observed colour.
For reference, we also illustrate the observed colours of a random subsample of spectroscopically confirmed QSOs from SDSS DR 14, as well as a set of confirmed stars that were identified independently of their observed photometric colours (or spectroscopic classification) - with robust ($>5\sigma$) measurements of parallax from \emph{Gaia} \citep[][see \citeauthor{2019AJ....157..168D}~\citeyear{2019AJ....157..168D} for details of cross matching to LS DR8 photometry]{2016A&A...595A...1G}.
Note that given the sensitivity limits of Gaia, parallax measurements are present only for bright stars within the LS DR8 catalogues ($m_{r}\lesssim 20$, corresponding to an average $13\%$ of PSF sources across the full catalogue).
Nevertheless, a stellar locus is clearly visible in each combination of observed colours, which is well traced by the region of high $P_{\text{star}}$.
The importance of WISE mid-IR information on robustly separating the star and QSO populations is also clearly demonstrated, with significant overlap between the two populations when only optical information is used ($g - r$ vs $r - z$), but a clear separation between the two loci when optical and mid-IR colours are combined.

\section{Photo-$z$ Methodology}\label{sec:method}
As with the previous studies upon which our methodology builds \citep{2019A&A...622A...3D,Duncan:21}, the primary algorithm we employ to derive photo-$z$ estimates is the sparse Gaussian process (GP) photometric redshift code, \textsc{GPz} \citep{2016MNRAS.455.2387A, 2016MNRAS.462..726A}, which extends the standard GP method to incorporate key features suited to photo-$z$ estimation.
Overall, \textsc{GPz} provides several benefits over standard GP implementations, including the use sparse of GPs that reduces computational requirements, with no loss in accuracy of the resulting models.
Additionally, \textsc{GPz} accounts for non-uniform and variable noise (heteroscedastic) within the input data - modelling both the intrinsic noise within the photometric data and model uncertainties due to limited training data.
Finally, \textsc{GPz} also allows for cost-sensitive learning (CSL), allowing different parts of parameter space to be weighted more or less importantly based on the specific scientific requirements.

In a key change from previous studies where we applied \textsc{GPz} to a range of datasets \citep{Duncan:2017ul,2019A&A...622A...3D,Duncan:21}, in this analysis we employ a new C++ implementation of \textsc{GPz}, \texttt{gpz++}\footnote{\url{https://github.com/cschreib/gpzpp}}, which provides substantial speed improvements and also implements support for missing data.
Furthermore, \citet{2020MNRAS.498.5498H} recently outlined a number of key augmentations to the photo-$z$ approach, which together lead to significant improvements in the resulting estimates compared to more straightforward applications of \textsc{GPz}. 
In the following subsections, we present details of our implementation for a number of these photo-$z$ training augmentations and other additional steps in our training methodology, before summarising the final training and prediction pipeline in Section~\ref{sec:final-method}.

\subsection{Population division with GMM}\label{sec:method-divide}
Amongst the approaches suggested by \citet{2020MNRAS.498.5498H}, one of the most effective additions was found to be the use of GMM to divide the population into $N$ distinct regions of colour-magnitude space, with \textsc{GPz} trained separately on each mixture.
In addition to providing improved estimates overall, this `GMM-Divide' technique was also found to result in improved training and prediction speeds due to the order of complexity intrinsic to the sparse GP algorithm.
In Section~\ref{sec:star-gal}, we found that modelling the population in 4 colour dimensions provides sufficient information to separate key populations and model the complex distribution of the optical samples.
However, the optimal way to divide the galaxy population for the purposes of photo-$z$ estimation represents a different and non-trivial problem, with even the limited available photometric bands available in LS DR8 providing an extremely complex high-dimensional parameter space.
This available parameter space increases when morphological information and apparent magnitudes (in addition to colour) are included, with both potentially incorporating more redshift information than a given colour.
Similarly, in addition to selecting the parameter space used to divide the population, when generating a GMM one must also define the number of Gaussian model components.

Unlike in Section~\ref{sec:star-gal}, when dividing the population for photo-$z$ training purposes we must also balance the accuracy of modelling the populations within the chosen parameter space with minimising the potential for empty or under-sampled training sets for individual populations (i.e. no spec-$z$ training sources belonging to mixtures defined from the full photometric sample).
Although the optimal number of components required to model the desired parameter space can be derived iteratively, i.e. choosing the number of components which minimises the Bayesian Information Criterion (BIC), this must be balanced against the size of the resulting training sample for each GMM component in the subsequent \textsc{GPz} estimation and the corresponding photo-$z$ performance.
Optimally modelling the observed colour space with the GMM therefore does not necessarily result in optimal photo-$z$ estimates for the available data. 

Based on initial tests, we find that including more than four dimensions within the GMM component results in unacceptable numbers of empty training sets for the current spectroscopic training samples.
For four GMM dimensions, the number of possible combinations for which data to use as input can not be practically be explored in full.
Our final approach to modelling the LS DR8 populations for the GMM-Divide methodology is therefore informed by prior knowledge on the most effective probes of redshift available within the optical catalogues, the distribution of expected signal to noise ratios for that observable and the results of manual experimentation.

First, the LS DR8 sample is separated into its individual morphological type (i.e. PSF, round exponential etc.), which implicitly separates the population into different parts of colour space.
For each morphological type, we then divide the population using GMMs constructed from the LS DR8 parameter space, with $g - r$ and $r - z$ colours and $m_z$ \emph{asinh} magnitudes used as the primary three dimensions for all sources.
For the resolved population (DEV, REX, EXP and COMP), the final GMM dimension was then selected to be the measured half-light radius, motivated by the positive impact of including size information in previous \textsc{GPz} photo-$z$ estimates \citep{2018MNRAS.475..331G}.\footnote{For the DEV and REX/EXP populations, we use the half-light radius of the deVaucouleurs (\texttt{SHAPEDEV\_R}) and exponential models respectively (\texttt{SHAPEEXP\_R}). For the composite population (COMP), which are modeled by the sum of deVaucouleurs and exponential models, we found that the half-light radii of the exponential components span a greater dynamic range and was therefore selected as the optimal size parameter for subsequent analysis.}
With size information unavailable for the unresolved (PSF) sources, $z-W1$ colour was chosen as the final GMM dimension due to its importance in separating stellar and extragalactic sources (Section~\ref{sec:star-gal}).
To ensure that the GMM is also representative of the \emph{extragalactic} PSF population for which we want to optimise photo-$z$ estimation, we only include PSF sources with $P_{\textup{star}} < 0.5$ and with parallax measurements $<3\sigma$ (if available).
\begin{figure}
\centering
 \includegraphics[width=1.02\columnwidth]{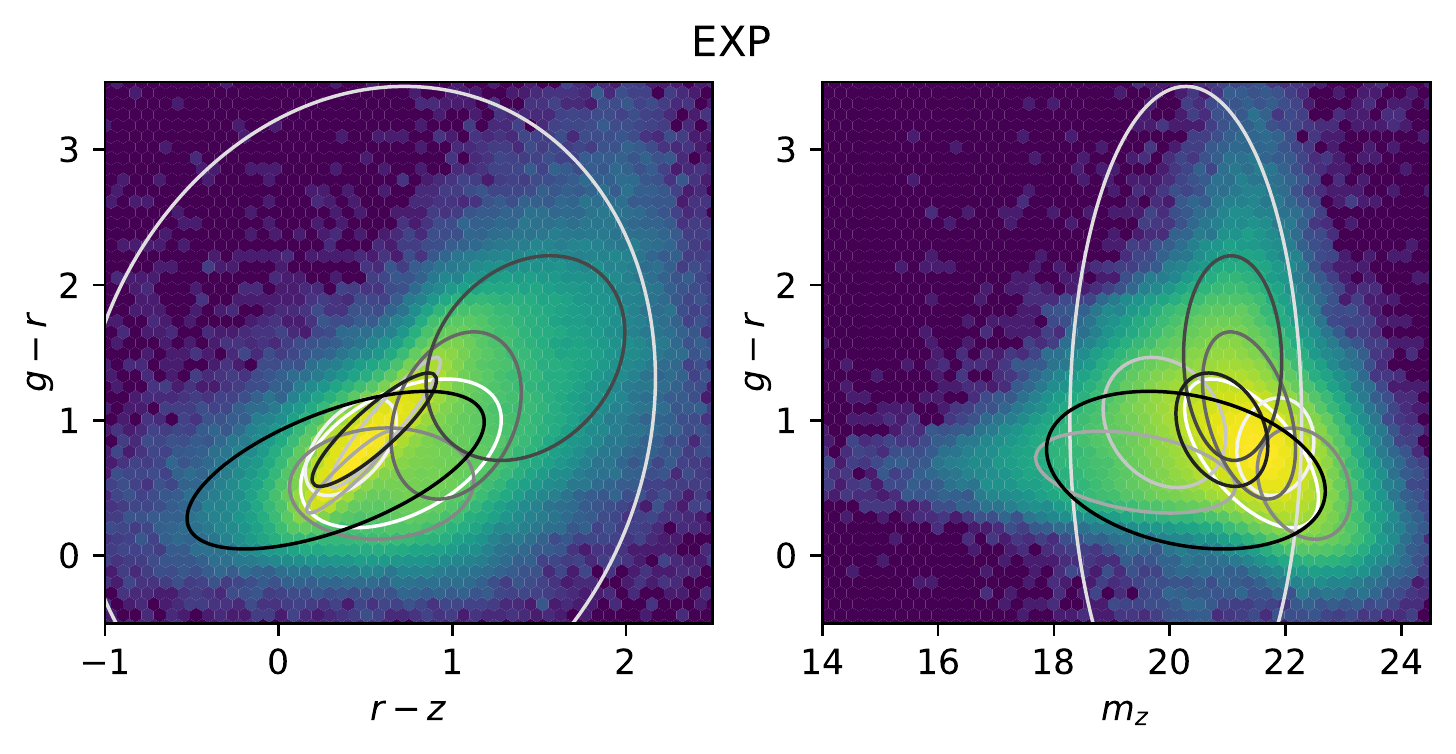}
 \includegraphics[width=1.02\columnwidth]{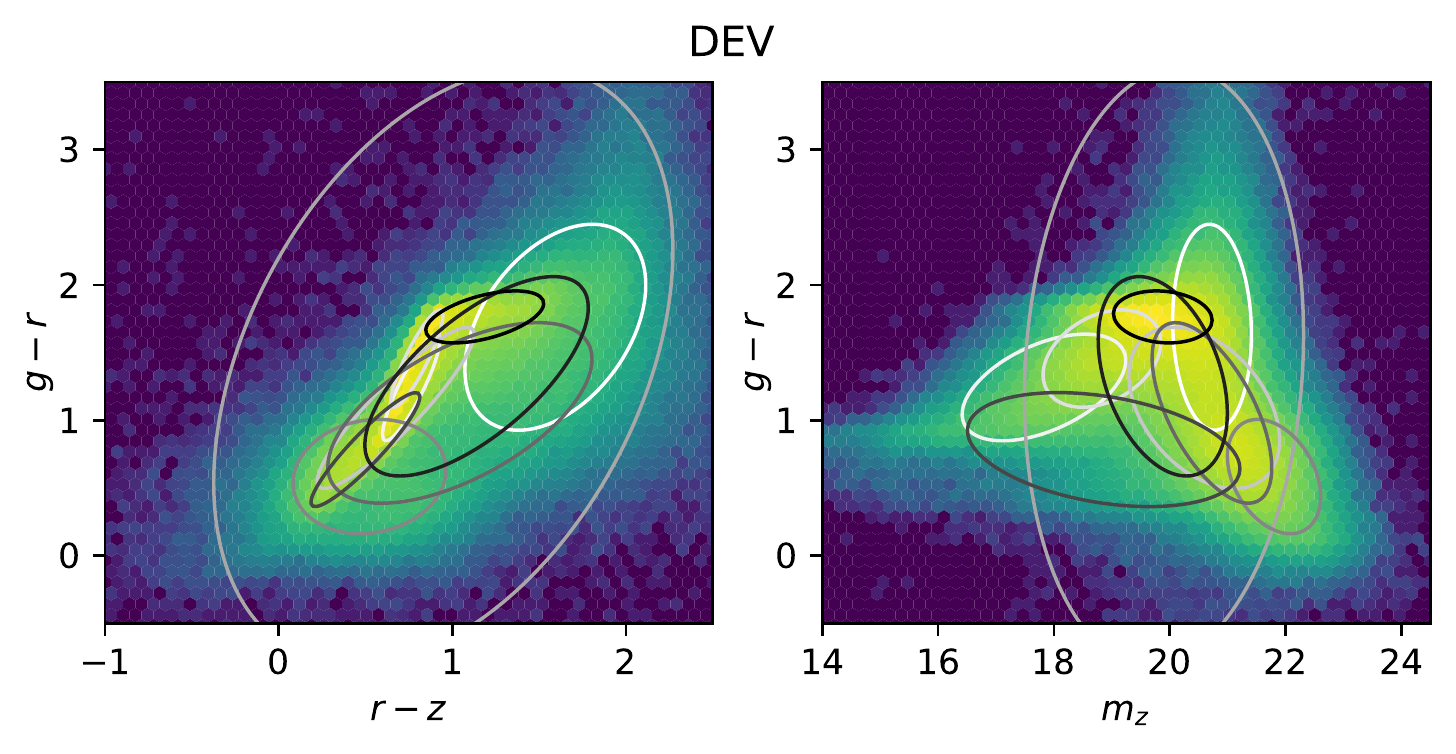}
 \caption{Optical colour-colour (left) and colour-magnitude (right) distributions of EXP and DEV morphologies within the Legacy dataset. The colour-scale shows the density of sources in each cell (logarithmically scaled). The means and covariances of the 10 components in the respective GMMs are over-plotted as ellipses.}
 \label{fig:gmm-divide}
\end{figure}

To illustrate how the observed parameter space can be modelled with the trained GMMs, in Fig.~\ref{fig:gmm-divide} we show two projections of full the colour-magnitude space for the EXP and DEV morphological populations.
Plotted over the density distributions for the full photometric sample are the means and covariances of the individual mixture components for the corresponding trained GMMs, where each GMM consists of 10 components.
In Section~\ref{sec:method-optimisation} we further explore how the number of GMM components impacts the overall quality of photo-$z$ predictions.

As outlined by \citet{2020MNRAS.498.5498H}, once the population has been modelled with a particular GMM, the subsequent population division can be incorporated into subsequent photo-$z$ analysis in a number of ways.
Firstly, sources can be assigned to their best-matching component based on which individual mixture model has the highest probability for the corresponding position in the 4-dimensional parameter space, with the photo-$z$ estimate taken from the \textsc{GPz} model trained on this component.
Alternatively, the photo-$z$ for a given source can be calculated as a weighted sum of the predicted photo-$z$s from the \textsc{GPz} models for all components, with the weighting derived from the relative probability of a source belonging to each component.
Unlike \citet{2020MNRAS.498.5498H}, who found that the majority of galaxies are assigned component probabilities near unity (and hence the weighted sum would produce identical results to the above approach), we find that a significant fraction of the catalogue sources have non-negligible probabilities of belonging to multiple GMM components.
The probabilistic approach could therefore potentially yield significant improvements in the overall photo-$z$ accuracy for the LS DR8 dataset.
However, due to the linearly increased computational requirements of predicting photo-$z$s for the full photometric sample ($k$ times longer prediction times for $k$ GMM components), such an approach was not deemed practical for the current analysis and the potential gains from using the linear superposition approach will be further explored in future studies.

Similar to the division of the full population, there is also flexibility in how the GMM is used to divide the spectroscopic training sample for training the respective instances of \textsc{GPz}.
As above, the simplest approach, and that adopted in this paper, is to assign each training source to its best-matching GMM component.
Alternatively, when training \textsc{GPz} for a given GMM component we can include all training sources with probability of belonging to the component greater than some non-negligible threshold, on the basis that including additional training sources on the periphery of the relevant parameter space should provide useful information.

\subsection{Weighted training samples for cost-sensitive learning}\label{sec:method-weight}
A second key approach suggested by \citet{2020MNRAS.498.5498H} employs the same GMM idea to calculate CSL weights for the training sample based on the relative probability density for a given set of features, $\mathbf{\hat{m}}$ (i.e. magnitudes and colours), in both the training sample and the full parent photometric sample.
Cost-sensitive learning allows for minimising potential biases in training from spectroscopic training samples that are not fully representative of the photometric sample.
For details of how CSL weights are incorporated into the objective function of \textsc{GPz} during training, we refer the reader to Section~4.2 of \citet{2016MNRAS.455.2387A}.

Similar to the nearest-neighbour based method employed by \citet{Lima:2008eu} and \cite{Duncan:2017ul}, the use of GMM to model the desired parameter space minimises the potentially negative impacts of small number statistics in some regions of parameter space.
Following \citet{2020MNRAS.498.5498H} we define the weighting for a given source as:
\begin{equation}
	w_{i} = \frac{p_{\textup{full}}(\mathbf{\hat{m}}_{i}) + \epsilon}{p_{\textup{train}}(\mathbf{\hat{m}}_{i}) + \epsilon},
\end{equation}
where $p_{\textup{full}}(\mathbf{\hat{m}}_{i})$ and $p_{\textup{train}}(\mathbf{\hat{m}}_{i})$ correspond to the probability density for a source with photometric properties, $\mathbf{\hat{m}}_{i}$, in the GMMs for the full photometric population and spectroscopic training sets respectively.
The additional constant, $\epsilon$, is included to prevent sparsely populated regions of parameter space (typically in the smaller training sample GMM) from producing extreme weights that negatively impact the photo-$z$ training. 
Due to the significantly increased training sample size in this work compared to that available in \citet{2020MNRAS.498.5498H}, we choose a smaller value of $\epsilon = 0.001$ (c.f. 0.01) which enables greater dynamic range in the resulting CSL weights.
Similarly, the maximum CSL weight allowed in our training was set to an increased value of 100 (c.f. 20).

\begin{figure}
\centering
 \includegraphics[width=1.0\columnwidth]{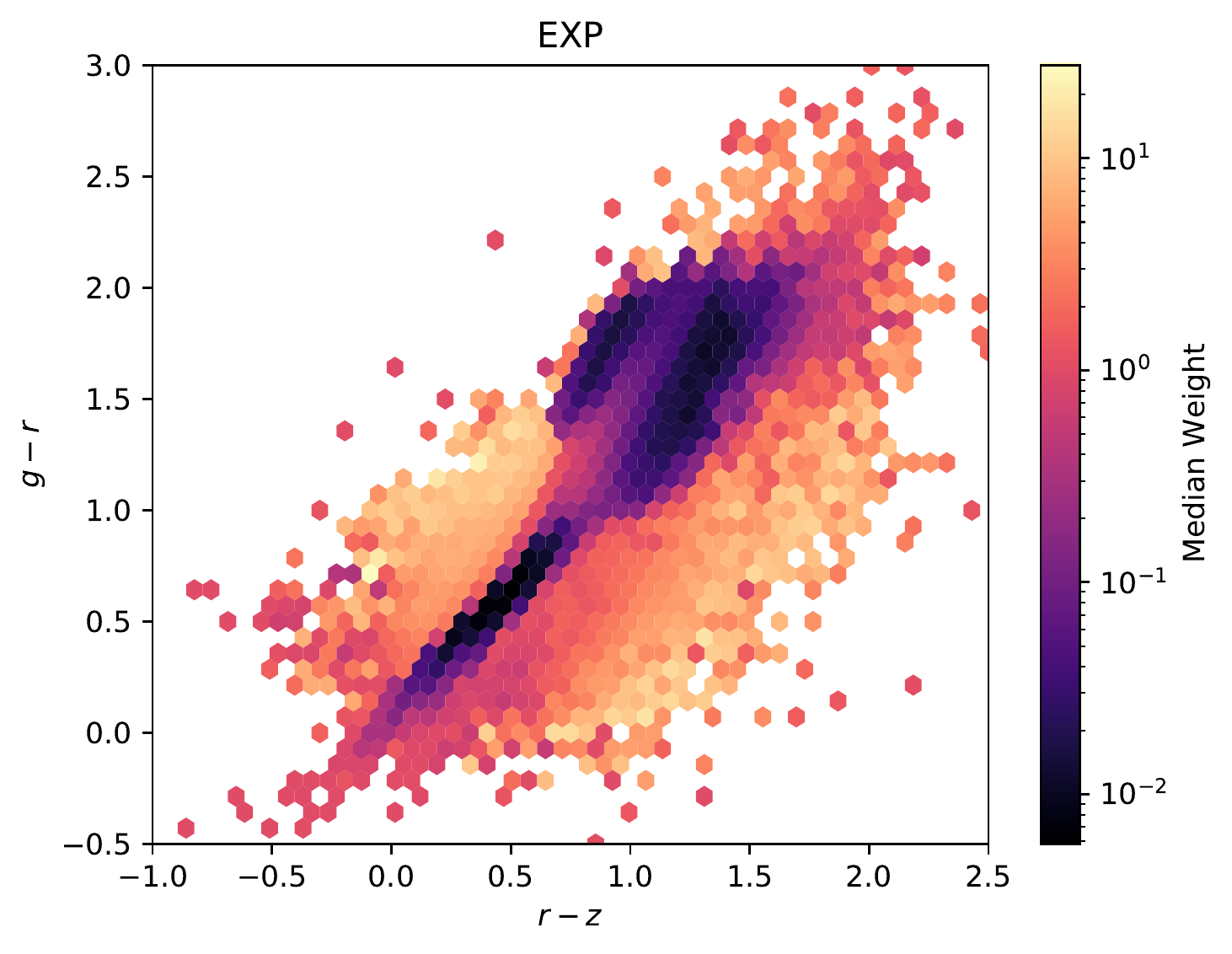}
 \includegraphics[width=1.0\columnwidth]{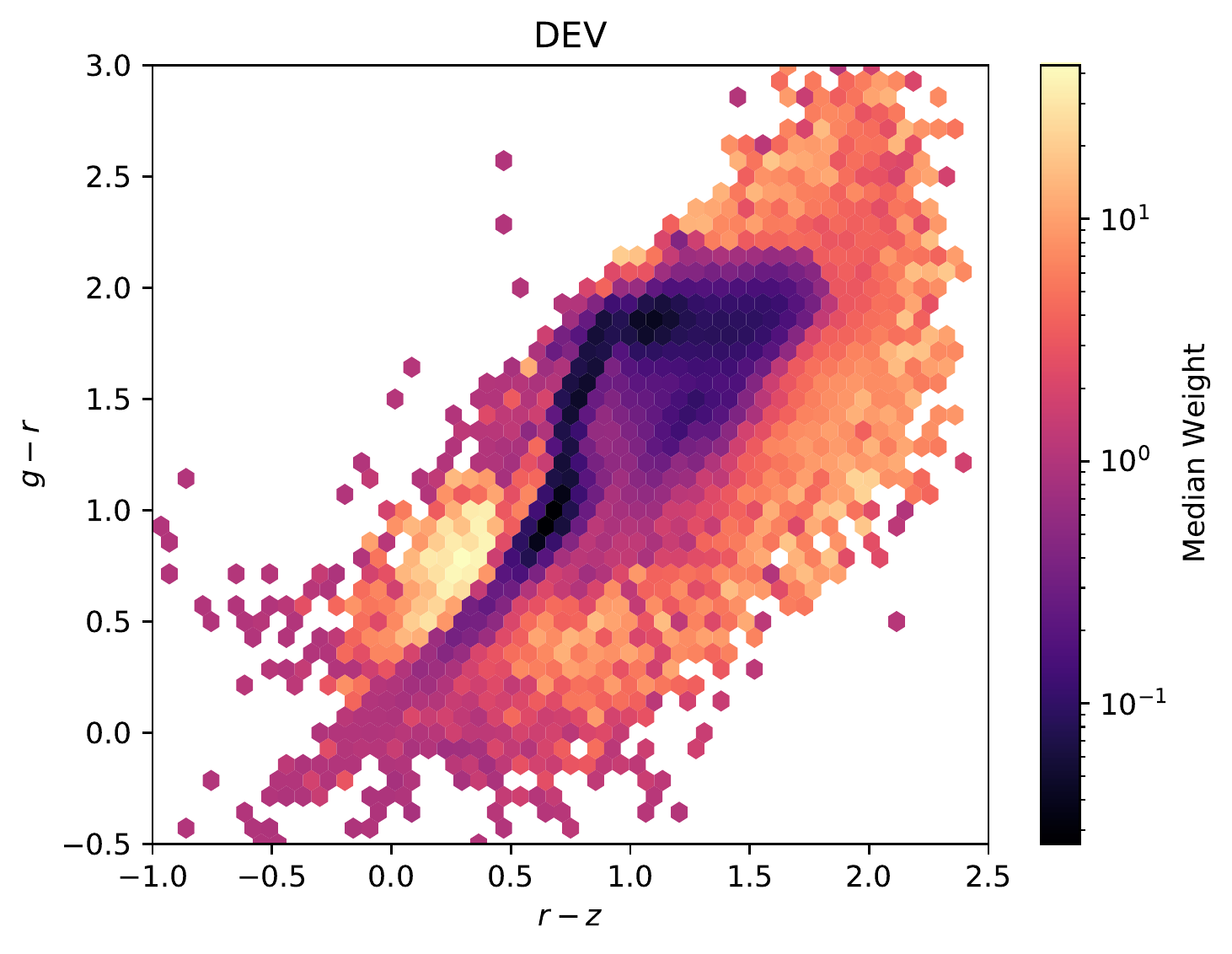}
 \caption{Illustration of the average CSL weight as a function of galaxy colour for two different morphologies in the photometric catalogues. The regions of dark colours (with medians weights $\lesssim10^{-1}$) highlight areas of parameter space with extensive training data such as the SDSS and BOSS/eBOSS ELG and LRG.}
 \label{fig:gmm-weight}
\end{figure}

For self-consistency and computational efficiency, we employ the same GMM models of the colour-magnitude magnitude space as generated in Section~\ref{sec:method-divide}, combining the $g - r$ and $r - z$ colours and $m_z$ \emph{asinh} magnitudes for all sources, with the addition of either $z-W1$ colours for unresolved (PSF) sources or the measured half-light radius for resolved sources (DEV, REX, EXP and COMP).
Fig.~\ref{fig:gmm-weight} illustrates the resulting distribution of weights derived for the EXP and DEV populations using this method, showing the average CSL weights for a 2D ($g - r$ versus $r - z$) projection of the respective colour-magnitude(-size) space.
Across $g - r$/$r - z$ colour space, the median weights span over three orders of magnitude, demonstrating the significant variation in spectroscopic training samples available over the parameter space probed by LS DR8 photometry.
Regions of parameter space which are significantly over-represented in the spectroscopic samples can be also clearly seen for both morphological classes, resulting in very low CSL weights.
These over-dense regions are a result of large colour-selected samples from SDSS BOSS and eBOSS surveys, such as the emission line galaxy (ELG) and luminous red galaxy (LRG) samples for the EXP and DEV morphologies respectively.

\subsection{Optimisation of photo-$z$ method hyper-parameters}\label{sec:method-optimisation}
As outlined in Section~\ref{sec:method-divide}, the precise impacts of how we choose to model (and hence divide) the population with GMMs on the resulting bulk photo-$z$ estimates represents a complex choice.
Furthermore, how best to combine the population division and CSL weighting across the full range of parameter space also presents an open question.
To quantitatively test the impacts of different methodology choices on the resulting final photo-$z$ estimates and find the optimal approach for our dataset (and scientific objectives), we therefore perform a systematic test of a range of approaches.
Specifically, we investigate two primary choices: the number of mixtures used when fitting GMMs to observational parameter space, and the inclusion (or not) of colour/magnitude/size dependent CSL weights during photo-$z$ training.

Using the full spectroscopic training sample for each morphological type, we perform the `GMM-Divide' step outlined in Section~\ref{sec:method-divide} and train the associated \textsc{GPz} photo-$z$ models for GMMs with a number of components, $N$.
For all iterations, the total number of basis functions used for training is 500 (with 500/$N$ used to train the \textsc{GPz} model for each GMM component).
As input properties for training \textsc{GPz}, we use the seven \emph{asinh} magnitudes available for all sources ($g$, $r$, $z$ \& WISE 1-4), with half-light radius also included for resolved sources.

We evaluate the resulting photo-$z$ performance by calculating bulk quality statistics for the spectroscopic test samples for each group, using the robust scatter, $\sigma_{\textup{NMAD}}$, and an absolute outlier fraction, $\text{OLF}_{0.15}$. 
Following common literature definitions \citep[e.g. ][]{Dahlen:2013eu}, we define 
\begin{equation}
\sigma_{\textup{NMAD}} =1.48 \times \text{median} ( \left | \delta z \right | / (1+z_{\textup{spec}})),
\end{equation}
where $\delta z = z_{\textup{phot}} - z_{\textup{spec}}$.
Similarly, we define outliers as sources where 
\begin{equation}
\left | \delta z \right | / (1+z_{\text{spec}}) > 0.15 .
\end{equation}

Since we are interested in only the relative performance of different approaches in this comparison, to minimise the impact of the least well constrained photo-$z$s on the overall statistics, for these comparison tests we exclude the 20\% of sources with the worst relative photo-$z$ uncertainty for each type ($\sigma_{z}/(1+z_{\text{phot}})$).
We then derive a balanced estimate of the photo-$z$ quality that is representative of the full population, calculating $\sigma_{\textup{NMAD}}$ and $\text{OLF}_{0.15}$ for 1000 sources that are drawn randomly (with replacement) from the respective test sample, with probabilities proportional to the CSL weight for each source.
We then repeat the random sampling 100 times and take the mean of the resulting distribution as the final estimate of $\sigma_{\textup{NMAD}}$/$\text{OLF}_{0.15}$ and the standard deviation of the distribution as an estimate of the corresponding uncertainty due to sample statistics.

\begin{figure}
\centering
 \includegraphics[width=0.98\columnwidth]{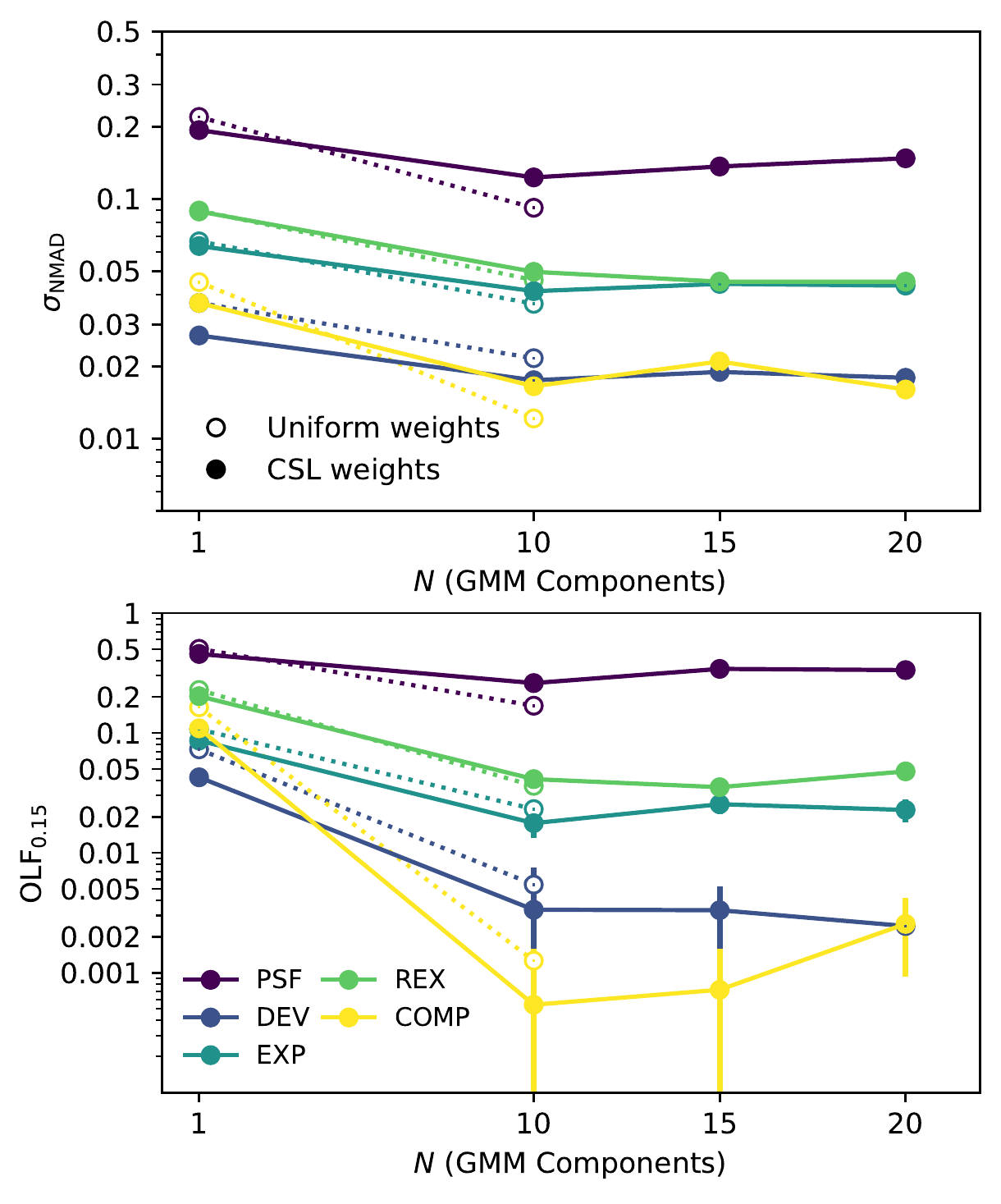}
 \caption{Representative photo-$z$ performance as a function of the number of components, $N$, used to divide the population with GMMs. For $N=1$, \textsc{GPz} is trained for the full population at once. Filled symbols (solid lines) show the results when CSL weights are included in the photo-$z$ training while open symbols (dotted lines) show the corresponding statistics for when \textsc{GPz} is trained with uniform weighting.}
 \label{fig:gmm_ncomp_stats}
\end{figure}

Figure~\ref{fig:gmm_ncomp_stats} presents the resulting photo-$z$ performance as a function of $N$ for each optical morphology sample (filled symbols and solid lines).
We find that for all morphologies, the resulting scatter and outlier fractions are reduced when the population is split using the GMM, with the most significant improvement seen in the resulting outlier fractions.
However, we see that once the population has been sufficiently divided, there is no universal improvement to be gained from more detailed modelling of the paramater space (i.e. larger $N$).
Results for some morphologies are further improved with more GMM components (e.g. improved scatter and outlier fraction for REX sources, or outlier fractions for DEV sources), but others deteriorate, most notably PSF and EXP sources.
When examining the performance for rarer populations of interest, we find that the conclusions drawn from the full test population are still valid.
For example, calculating the scatter and outlier fraction for $z > 3$ QSOs in the test sample (with no weighting included in the statistics) we find that using 10 GMM components results in substantially improved outlier fractions compared to training all PSF sources simultaneously, with $\text{OLF}_{0.15}$ reducing from a very poor 0.71 to $\text{OLF}_{0.15} = 0.2$ (with $\sigma_{\textup{NMAD}}$ improving from 0.12 to 0.08).
For 15 or 20 GMM components the $\text{OLF}_{0.15}$ worsens to $\approx 0.32$, confirming that for the current available training samples 10 GMM components represents an optimal choice over a wide range of parameter space, although all options with larger $N$ would yield comparable results.

The second key methodology decision we explore is the inclusion of CSL weights derived from the GMM distributions (Section~\ref{sec:method-weight}).
When no population division is employed (i.e. $N = 1$), the incorporation of CSL weights during training leads to improvement in robust scatter and outlier fraction for all source types (open symbols and dotted lines in Fig.~\ref{fig:gmm_ncomp_stats}).
However, when combining population division and CSL weights the results are more varied, with some morphologies seeing better performance (both scatter and outlier fractions) with CSL weights included and others performing better with uniform training weights.
Motivated by the improved overall performance (or negligible loss in performance) for the resolved optical morphologies when CSL weights are included alongside sample division, we opt to include CSL weights as standard for our final methodology.

Finally, as discussed in Section~\ref{sec:method-divide}, an additional variation that can be explored is how spectroscopic training sources are assigned to GMM components derived in the GMM-Divide step.
However, when training \textsc{GPz} including all training sources with $>10\%$ probability of belonging to a given mixture and training including sources with $>50\%$ probability, we find no significant difference in the resulting photo-$z$ statistics. 
For simplicity and speed, we therefore train \textsc{GPz} for GMM components using only the training samples which are best-matched to the given component.

In addition to the steps outlined in Sections~\ref{sec:method-divide} and \ref{sec:method-weight}, \citet{2020MNRAS.498.5498H} also found some improvement in accuracy of the photo-$z$ posterior predictions could be gained from training multiple iterations of \textsc{GPz} where the training data had been perturbed based on the photometric noise (resulting in non-Gaussian posteriors).
However, we determined that the small statistical improvement would not be of sufficient benefit to justify the linearly increased training and prediction times associated.

\begin{figure*}
\centering
 \includegraphics[width=1.0\textwidth]{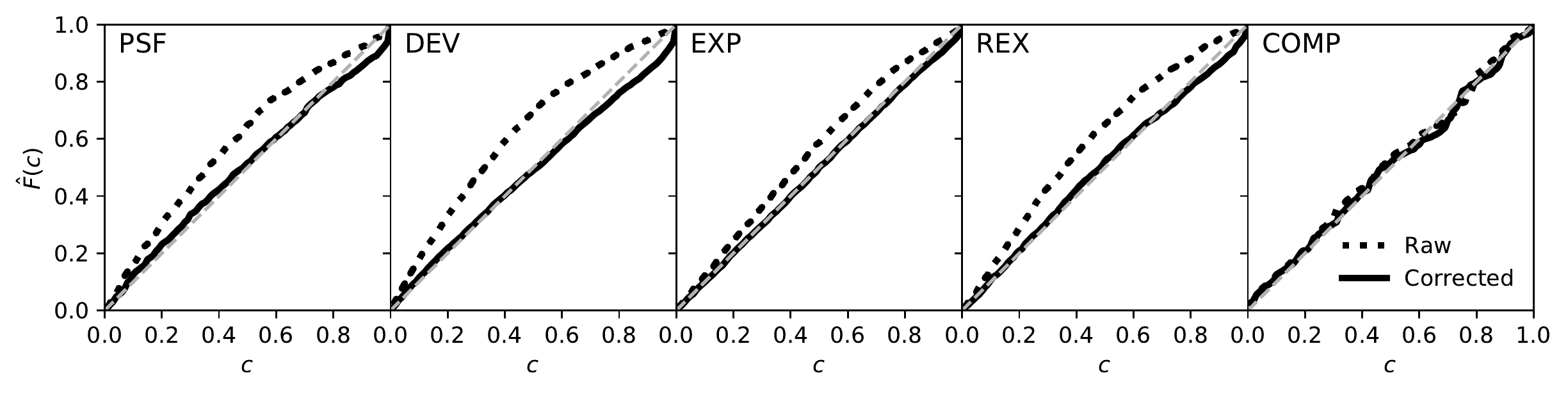}
 \caption{Panels illustrate the cumulative distribution of threshold credible intervals, $c$, ($\hat{F}(c)$) for the spectroscopic test sample before (dotted lines) and after (solid lines) uncertainty calibration. The cumulative distribution is weighted using the CSL weights to provide a representative view of the overall uncertainties. Lines that rise above the 1:1 relation illustrate under-confidence in the photo-$z$ uncertainties (uncertainties overestimated) while lines that fall under illustrate over-confidence (uncertainties underestimated).}
 \label{fig:error_calibration}
\end{figure*}

\subsection{Calibration of photo-$z$ uncertainties}\label{sec:method-errorcalib}
One of the key strengths of the \textsc{GPz} photo-$z$ algorithm is its ability to predict accurate uncertainties, with the accuracy of the uncertainty estimates fully incorporated into the objective function during model training. 
However, due to the fact that the spectroscopic training sets cannot be fully representative (and CSL weights cannot perfectly account for these biases), input photometric uncertainties may be over or under-estimated, and its photo-$z$ posteriors are not multi-modal, previous studies have found that additional calibration of the uncertainties after training is still necessary \citep{Duncan:2017ul,2019A&A...622A...3D}. 
We therefore include an additional post-training error calibration step in our method.

To quantify the over- or under-confidence of our photo-$z$ estimates, we follow the method outlined in Section~3.3.1 of \citet[][itself based on the proposal by \citeauthor{2016MNRAS.457.4005W} \citeyear{2016MNRAS.457.4005W}]{Duncan:2017ul}, whereby we calculate the distribution of threshold credible intervals, $c$ (where the spectroscopic redshift intersects the redshift posterior).
For perfectly accurate uncertainty estimates, the cumulative distribution of credible intervals, $\hat{F}(c)$, should follow a straight 1:1 relation, i.e. a quantile-quantile (or $Q-Q$) plot.
Curves that fall below this 1:1 relation indicate that the photo-$z$ errors are underestimated (i.e. the $P(z)$s are too sharply peaked), while curves that rise above indicate uncertainties are over-estimated and the predictions are under-confident.
As the uncertainties predicted by \textsc{GPz} are Gaussian, we calculate the threshold credible interval for a given source, $c_{i}$, following 
\begin{equation}
c_{i} = \textup{erf} \left (\frac{| z_{i,\textup{spec}} - z_{i,\textup{phot}} |}{\sqrt{2} \sigma_{i}} \right),	
\end{equation}
where $z_{i,\textup{spec}}$ and $z_{i,\textup{phot}}$ are the true (spectroscopic) and predicted redshift estimates and $ \sigma_{i}$ the corresponding predicted 1-$\sigma$ photo-$z$ uncertainty.
For the majority of source morphologies and apparent magnitudes in our trained \textsc{GPz} models (independent of the precise augmentation steps applied), we find that the photo-$z$ uncertainties are typically over-estimated.
When calibrating the uncertainties, we scale the original uncertainties as a function of magnitude, $m$, following
\begin{equation}\label{eq:gpz_err_scale}
	\sigma_{\textup{new},i} = \sigma_{\textup{old},i} \times \alpha(m_{i}),
\end{equation}

\noindent where $\alpha(m)$ is defined as
\begin{equation}\label{eq:smoothing_2}
	\alpha(m) = \begin{cases}
	 \alpha_{\eta} & m \leq m_{\eta}\\
	 \alpha_{\eta} + \kappa \times(m-m_{\eta}) & m > m_{\eta},
	\end{cases}
\end{equation}
\noindent with $\alpha(m)$ constant ($\alpha_{\eta}$) below a chosen characteristic apparent magnitude, $m_{\eta}$, and scaling linearly with magnitude for sources fainter than $\alpha_{\eta}$ \citep{2009ApJ...690.1236I}.
We use the $r$-band optical \emph{asinh} magnitudes for calculating the magnitude dependence of the error scaling and assume a characteristic magnitude of $m_{\eta} = 16$.
The parameters $\alpha_{\eta}$ and $\kappa$ are then fitted for each morphology type individually (combining all respective GMM components) using the \textsc{emcee} Markov chain Monte Carlo fitting tool \citep[MCMC;][]{2013PASP..125..306F} to minimise the Euclidean distance between the measured and ideal $\hat{F}(c)$ distributions.
In this work, we include the additional step of weighting the measured $\hat{F}(c)$ distribution using the CSL weights derived in Section~\ref{sec:method-weight} to mitigate for biases in the spectroscopic training and validation samples used to calibrate photo-$z$ uncertainties.

Fig.~\ref{fig:error_calibration} illustrates the accuracy of the photo-$z$ uncertainty estimates of one set of photo-$z$ estimates (10 GMM components, CSL weights included; see below) before and after uncertainty calibration.
In Fig.~\ref{fig:error_calibration} we present an estimate of the overall photo-$z$ uncertainty based on the spectroscopic test sample weighted by the CSL weights derived in Section~\ref{sec:method-weight}.
A general trend visible in our results is that prior to the additional calibration, the photo-$z$ uncertainties predicted by \texttt{gpz++} are under-confident, overestimating the uncertainties across a wide range of magnitudes and across all morphologies.
Following the calibration step, the overall weighted uncertainty distributions are significantly improved, very closely following the optimal 1:1 relation.
However, for some morphologies (PSF, DEV, REX) the $\hat{F}(c)$ distribution falls slightly below the 1:1 relation at credible intervals of $0.8 < c < 1$, indicating that the tails of the photo-$z$ posteriors are underestimated.
This effect illustrates a limitation of the assumption of purely Gaussian posteriors in our analysis, which additional \textsc{GPz} augmentation steps could further mitigate in future.
Nevertheless, Fig.~\ref{fig:error_calibration} demonstrates that the overall accuracy of the corrected  final photo-$z$ uncertainties are extremely accurate.

\subsection{Final photo-$z$ training and prediction pipeline}\label{sec:final-method}
Based upon the systematic tests outlined in Section~\ref{sec:method-optimisation}, the final photo-$z$ training methodology for our analysis is as follows:
\noindent the training sample and full population for each morphological type is modelled in four dimensions ($g - r$, $r - z$, $m_z$ and $z-W1$ or half-light radius) using 10 GMM components.
A total of 500 basis functions for each morphological type are then used to train \textsc{GPz} (50 per mixture model), with CSL weights included in the training for all components.
As above, we use the seven \emph{asinh} magnitudes available for all sources ($g$, $r$, $z$ \& WISE 1-4), plus the half-light radius for resolved sources, as inputs for training and prediction.
Sources within the training sample are used to train only their best-matching mixture model.
For all \textsc{GPz} training steps, the training sample is split such that 60\% is used for training and 20\% used for validation during training (to prevent over-fitting). 
The remaining 20\% is retained for testing and quantitative analysis in Section~\ref{sec:results}.

For the final production run, we train \textsc{GPz} models for each GMM component in two stages, with the first pass trained using only the simpler variable diagonal covariances (GPVD) between parameters.
The trained GPVD model for each GMM component is then used as a starting point for a second training iteration using the full variable covariances (GPVC).
The two-stage training process allows us to make full use of the complex covariance modelling implemented by \textsc{GPz}, improving the overall accuracy of the photo-$z$ estimates and the estimated uncertainties (with smaller adjustments required in the uncertainty calibration stage), while also avoiding local minima during training and poor convergence during optimisation.
Due to the reduced complexity of the GP optimisation for the GPVD covariances, the increase in model training time is negligible. 
Overall, this two-stage training process generally results in improved efficiency due to the prevention of training failures resulting from \textsc{GPz} models not reaching convergence. 
The final trained GMM for each morphology and the full set of associated \texttt{gpz++} models (one for each GMM component) are stored for future predictions. 

\begin{figure*}
\centering
 \includegraphics[width=0.9\textwidth]{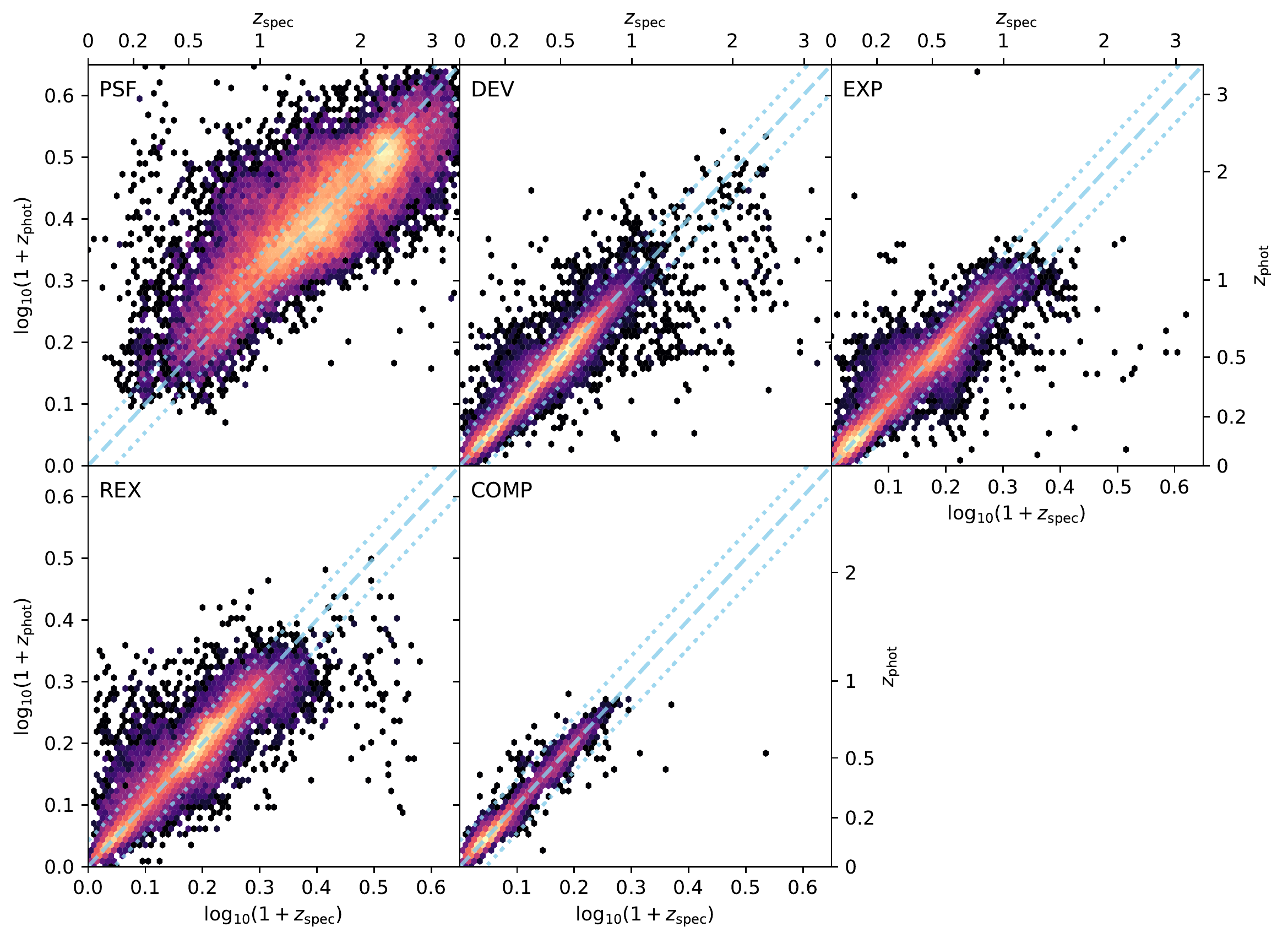}
 \caption{Distribution of estimated photo-$z$ as a function of true spectroscopic redshift for each morphological type in the spectroscopic test sample (i.e. sources not included in any training or validation stage). Only sources with well-constrained estimates ($\sigma_{z} / (1+z_{\text{phot}}) < 0.2$) are included in the plot. Colour scales illustrate the total number of sources in each cell with a logarithmic colour scale and the blue dashed and dotted lines illustrate the 1:1 and $\pm 0.1\times(1+z_{\text{spec}})$ relations respectively.}
 \label{fig:gpz_specz_photz}
\end{figure*}

For subsequent photo-$z$ prediction of new galaxy samples from LS DR8, we implement a pipeline which takes as input any catalogues in the format of the Legacy Surveys \texttt{tractor} catalogue. 
The pipeline performs all the steps required to produce photo-$z$ predictions, including: 
\begin{enumerate}
    \item Deriving \emph{asinh} magnitudes and reformatting of available size information.
    \item Homogenisation of photometry for LS DR 8 North sources (if required). 
    \item Calculation of star-QSO probability for PSF sources. Note that photo-$z$ predictions are derived for all PSF sources regardless of $P_{\text{star}}$, with the associated uncertainties and $P_{\text{star}}$ used to inform a final star-QSO classification alongside the photo-$z$ uncertainties. 
    \item Calculation of GMM component membership and CSL weights based on the source properties for the appropriate morphology (Sections~\ref{sec:method-divide} and ~\ref{sec:method-weight}).
    \item Production of \texttt{gpz++} input catalogues and prediction with the associated trained \texttt{gpz++} model for the corresponding GMM component.
    \item Adjustment of the output photo-$z$ uncertainties as a function of apparent magnitude and morphological type (Section~\ref{sec:method-errorcalib}).
    \item Combination and formatting of all separate component and morphology catalogues into a single output catalogue for a given input catalogue.
\end{enumerate}
By design, the full pipeline is ambivalent to the specific choices made regarding the  photo-$z$ training (i.e. the number of GMM components, inclusion of CSL weights, threshold for training sample model membership), accounting for all necessary book-keeping and merging. 
When used to predict photo-$z$ estimates for the full LS DR8 catalogue in bulk, the processing is split into individual \textsc{HEALPix} order 3 regions.\footnote{We note that when using 10 cores (20 threads) on an Intel Xeon W-2175 14C 2.5GHz processor, the full prediction pipeline is able to process $\sim3\times 10^{6}$ sources per hour.}
In the following section we present a detailed quantitative analysis of the resulting photo-$z$ estimates for the spectroscopic test sample, including comparison to existing estimates in the literature and investigation of the photo-$z$ performance of individual sub-populations of interest.

\section{Photo-$z$ Quality}\label{sec:results}

The goal of this work is to produce photo-$z$ estimates that perform well over the full range of source types and redshifts present in the LS DR8 photometric catalogues.
Fig.~\ref{fig:gpz_specz_photz} presents a qualitative illustration that demonstrates how well the full \textsc{GPz} methodology is able to achieve this goal, showing the distribution of photo-$z$ estimates as a function of spec-$z$ for the spectroscopic test sample where the photo-$z$ is deemed to be relatively reliable, i.e. that the uncertainty on individual estimates are small.
Specifically, here and in the remainder of this work we define a permissive criteria that photo-$z$ estimates are \emph{well-constrained} if
\begin{equation}\label{eq:err_cut}
\sigma_{z} / (1+z_{\text{phot}}) < 0.2.
\end{equation}
For all resolved morphologies, the photo-$z$ predictions are extremely precise and reliable at $z < 1$, with no evidence of systematic biases and relatively few catastrophic outlier.
Although exhibiting visibly increased scatter, robust estimates for PSF sources (predominantly QSOs) span $0.5 < z < 3$ and beyond.

\subsection{Overall photo-$z$ quality statistics}\label{sec:stats_overall}
To provide a more quantitative analysis of the photo-$z$ performance and evaluate the range of parameter space for which they can be used for different scientific analyses, it is important to characterise the statistical properties of the photo-$z$ estimates. 
In addition to calculating the $\sigma_{\text{NMAD}}$ and $\text{OLF}_{0.15}$ outlined above, in the following section we also examine the photo-$z$ performance including two additional statistics: the relative outlier fraction, $\text{OLF}_{3\sigma}$, defined as $\left | \delta z \right | / (1+z_{\text{spec}}) > 3\sigma_{\text{NMAD}}$, and the bias, $\Delta_{z} = \text{median} ( \delta z / (1+z_{\text{spec}}) )$.

\begin{figure}
\centering
 \includegraphics[width=1.0\columnwidth]{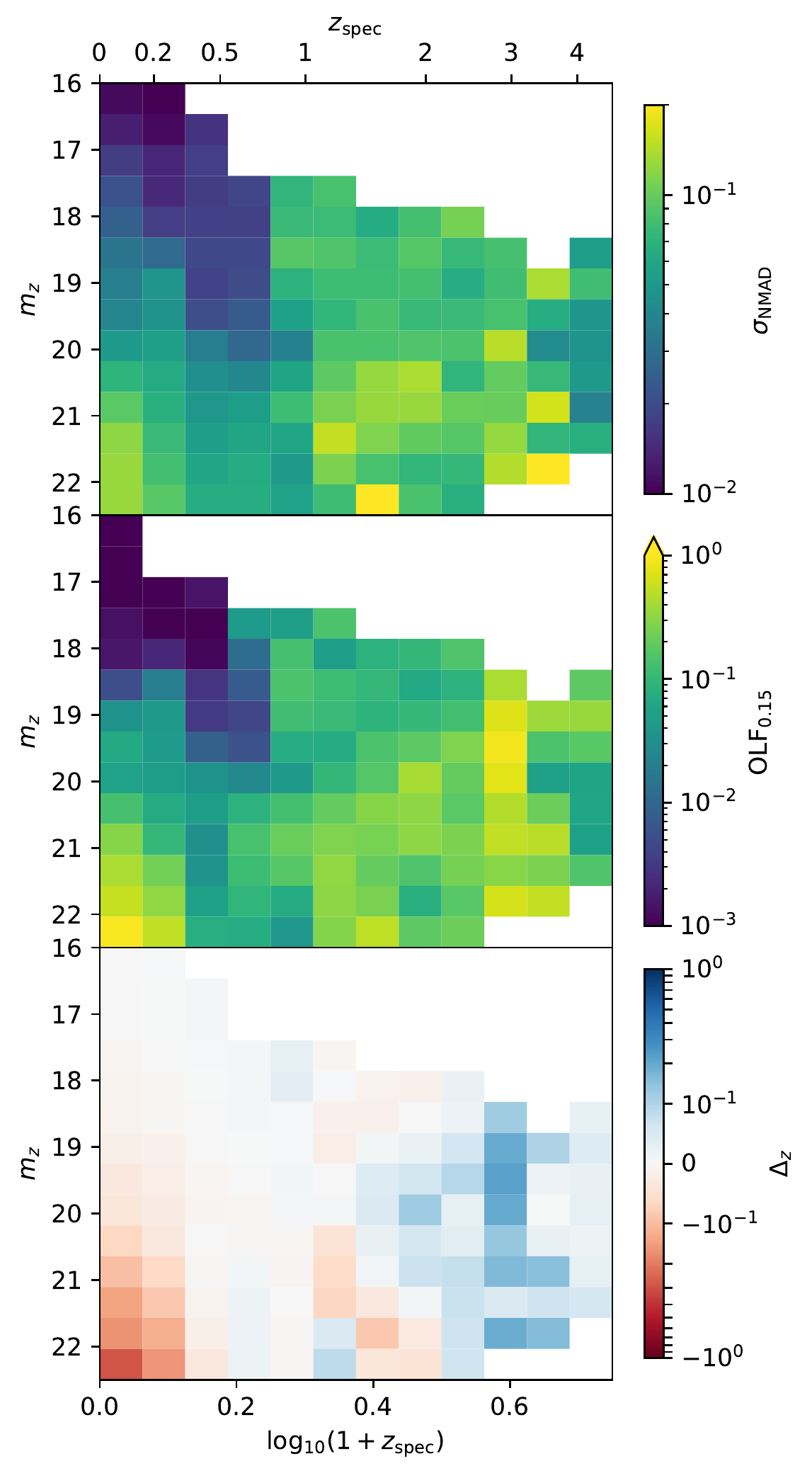}
 \caption{Robust scatter ($\sigma_{\textup{NMAD}}$; top), outlier fraction ($\textup{OLF}$; middle) and bias ($\Delta_{z}$; bottom) as a function of spectroscopic redshift and $z$-band magnitude for the full spectroscopic test sample. For a cell to be plotted we require a minimum of five galaxies.}
 \label{fig:gpz_stats_z_mag}
\end{figure}

In Fig.~\ref{fig:gpz_stats_z_mag} we show how $\sigma_{\text{NMAD}}$, $\text{OLF}_{0.15}$ and $\Delta_{z}$ vary as a function of true redshift and apparent magnitude, $m_{z}$, for the full spectroscopic training sample (including all morphologies). 
As in Fig.~\ref{fig:gpz_specz_photz}, analysis is restricted to photo-$z$ that are estimated to be well-constrained (satisfying Eq.~\ref{eq:err_cut}) and free from significant source blending or optical artefacts.
We find that both $\sigma_{\text{NMAD}}$ and $\text{OLF}_{0.15}$ follow a similar trend, with average photo-$z$ quality being a strong function of redshift and magnitude - typically with accuracy, precision and outlier fraction deteriorating for fainter magnitudes and the highest redshifts as expected.
The increase in scatter and outlier fraction towards fainter magnitudes is most significant at $z_{\text{spec}} < 1$, where the LS DR8 population is dominated by resolved morphologies.
At $z > 1$, where the catalogue becomes dominated by luminous QSOs, the overall photo-$z$ precision and reliability is reduced compared to low redshift but remains relatively consistent over a very broad range in redshift and magnitude.

When examining the bias, we find that the bias remains negligible across the majority of the redshift-magnitude space probed by LS DR8.
However, two distinct areas of parameter space exhibit significant bias: very faint sources at $z < 0.3$ which have photo-$z$ estimates that are significantly over-estimated, and QSOs at $2.5 \lesssim z \lesssim 4$ that are biased towards low redshifts.
In both cases, we postulate that the poorer performance in these regions of parameter space is primarily driven by the limitations in the available photometric information, rather than a failure of the photo-$z$ methodology itself or lack of training data. 
In particular, for faint sources where no size or mid-IR information are available, the lack of $u$-band photometry in LS DR8 means that we are unable to probe key features such as the Balmer or Lyman break \citep[see e.g. Fig.~2 and 3 of][for a systematic test on the impact of $u$-band photometry on photo-$z$ estimates]{2008MNRAS.387..969A}. 
Additionally, given the physical lower limit on redshifts within the training sample ($z=0$, excluding proper motion) and the overall redshift distribution of sources with $m_{z} > 20$, predictions for faint low-redshift sources that are poorly constrained will naturally be biased towards the higher redshifts at which the majority of the training sources with comparable magnitudes reside.

\subsection{Comparison with existing Legacy DR8 Photo-$z$s}\label{sec:zhou_comp}
As outlined in Section~\ref{sec:intro}, photo-$z$ estimates for the full LS DR8 catalogues do already exist within the literature.
The most recent and extensive of these estimates, presented by \citet{2021MNRAS.501.3309Z}, employs a random-forest machine learning approach applied to a combination of colour, magnitude and shape information based on the LS catalogues. 
To enable a direct comparison with the estimates from \citet{2021MNRAS.501.3309Z}, we extract the photo-$z$ predictions from both datasets within two regions of the sky with dense spec-$z$ sampling coverage down to magnitude limits sufficient to provide a representative sample.
Within the LS DR8 North footprint, we use the region around the NOAO Deep Wide Field Survey \citep[NDWFS][]{{Jannuzi:1999wu}} in Bo\"{o}tes, which incorporates spectroscopic surveys such as the AGN and Galaxy Evolution Survey \citep[AGES;][]{Kochanek:jy}.
In LS DR8 South we use the region around the XMM-LSS deep field, which contains numerous redshift surveys across a broad range of parameter space. 

After matching both sets of photo-$z$ estimates to the combined spectroscopic sample, we restrict the sample to only those sources which are free from source blending, are unaffected by masked regions in the LS DR8 imaging and have low probability of being a star ($P_{\text{star}} < 0.2$).
We then calculate the photo-$z$ quality statistics for both sets of estimates as a function of spec-$z$ using identical samples.
To estimate the potential variation in statistics given the smaller samples used in this comparison, we calculate the quality statistics for 500 bootstrap samples of a maximum of 100 sources in each bin, taking the median and 16/84th percentiles of the resulting distribution as our estimate and corresponding uncertainties.

\begin{figure}
\centering
 \includegraphics[width=0.95\columnwidth]{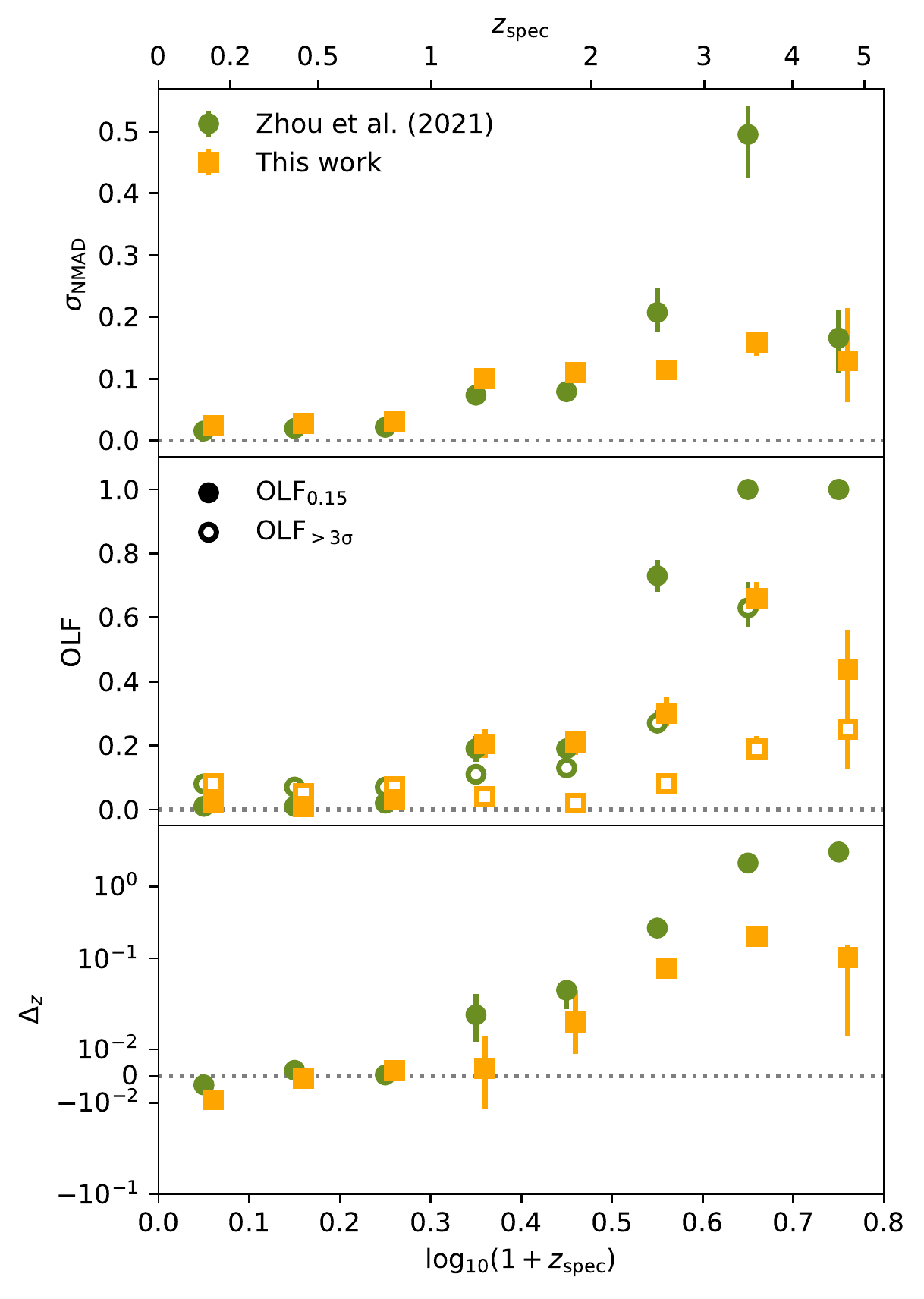}
 \caption{Photo-$z$ performance metrics as a function for spec-$z$ for the estimates of \citet[][green circles]{2021MNRAS.501.3309Z} compared to those of this work (gold squares). Figure panels show $\sigma_{\text{NMAD}}$ (top), $\text{OLF}_{0.15}$/$\text{OLF}_{3\sigma}$ (middle) and $\Delta_{z}$ (bottom). Error bars encompass the 16 to 84th percentiles of 500 bootstrap samples, each containing a maximum of 100 sources. For the definition of $\Delta_{z}$ used in this work, positive values mean that photo-$z$s are biased to lower redshifts than the true redshift.}
 \label{fig:zhou_stats_comparison}
\end{figure}

In Fig.\ref{fig:zhou_stats_comparison} we show the evolution of the key photo-$z$ statistics for both estimates as a function of spectroscopic redshift.
Overall, both sets of estimates provide excellent results at $z < 1$, with low scatter, outlier fraction and negligible bias.
We note that the estimates of \citet{2021MNRAS.501.3309Z} produce excellent results even beyond the parameter space for which they were designed.
However, at the highest redshifts where the LS DR8 photometric sample becomes dominated by QSOs, our \textsc{GPz} estimates perform significantly better, with substantially reduced bias (up to $>10\times$ lower)  and absolute outlier fraction ($\text{OLF}_{0.15}$; $2-3\times$ lower).
At $1 < z < 2$, where LS DR8 contains a mix of luminous galaxies and QSOs, the photo-$z$ estimates in this work are significantly less biased and have substantially reduced fraction of catastrophic outliers ($\text{OLF}_{3\sigma}$) at the expense of marginally increased scatter.
 These overall statistics demonstrate the difference in the respective intended use case and methodologies, with the photo-$z$ estimates produced here trading a small amount of performance for bright and well sampled populations (e.g. $z < 1$ and $m_{z} < 21$) for much more robust estimates of rarer or faint populations in LS DR8.
    
For studies at low redshift, where the lowest possible scatter and bias for a magnitude selected and resolved population is essential (e.g. clustering analysis of luminous red galaxies), the \citet{2021MNRAS.501.3309Z} therefore offer some increased performance compared to those in this work.
 However, due to the significant bias observed for high-$z$ sources (incorrectly estimated to be $z < 1$), purely photo-$z$ selected samples could be contaminated by interlopers.
 For studies that require reliable photo-$z$ estimates for the widest range of potential source types, the photo-$z$ estimates provided here should therefore provide substantially more reliable and less biased results - with negligible loss of precision for $z < 1$ population.

\subsection{Performance for rare or extreme populations}
In the following section, we explore the performance of our photo-$z$ estimates for a number of rare but scientifically import subsets of the LS DR8 photometric catalogues.

\subsubsection{High redshift quasars}\label{sec:results_highz}
\begin{figure}
\centering
 \includegraphics[width=0.92\columnwidth]{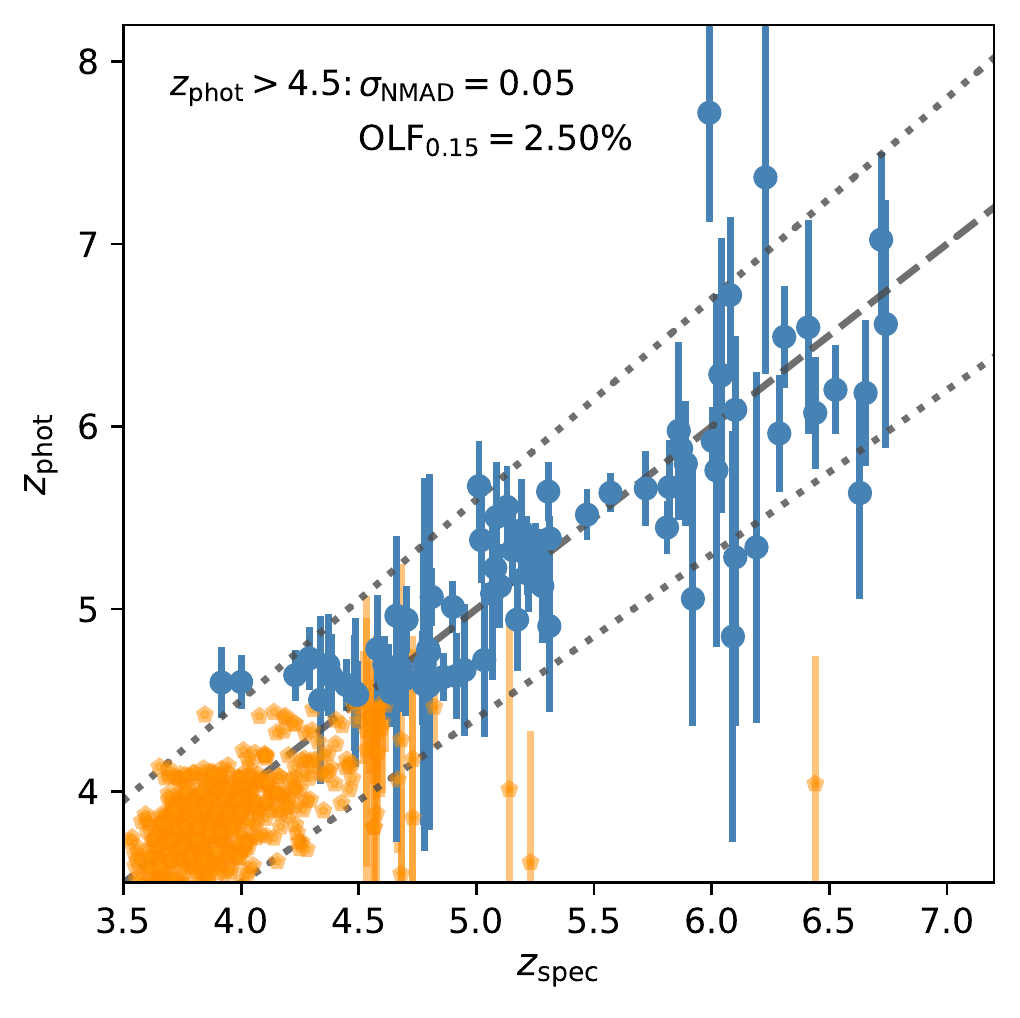}
 \caption{Comparison of predicted, $z_{\text{phot}}$, vs true redshift, $z_{\text{spec}}$, for the highest redshifts probed by the spectroscopic test sample. Sources selected to have $z_{\text{phot}} > 4.5$ with well-constrained estimates (Eq.~\ref{eq:err_cut}) are plotted as solid blue circles with associated 1$\sigma$ uncertainties. Additional spectroscopic test sample sources with $z_{\text{spec}} > 3.5$ are shown as orange pentagons (1$\sigma$ uncertainties are also shown for $z_{\text{spec}} > 4.5$ that are not selected through the photo-$z$ criteria).}
 \label{fig:hzq}
\end{figure}

As partially illustrated in Section~\ref{sec:zhou_comp}, a common problem for many empirical photo-$z$ estimates derived for samples that span a wide range of source properties and redshifts is that predictions for the highest redshift sources can become extremely biased towards low redshift.
These biases can be present even when the photometry involved is able to probe key redshift features such as the Lyman break and training samples extend beyond the relevant redshift range \citep[see e.g.][for examples across multiple datasets and machine learning algorithms]{2019PASP..131j8004N,2020A&A...644A..31E}.
To verify the performance of our photo-$z$ estimates out to the highest redshifts probed by the LS DR8 optical data, we select all sources from the spectroscopic redshift sample retained for testing that have both a predicted redshift of $z_{\text{phot}} > 4.5$ and have uncertainties that satisfy Eq.~\ref{eq:err_cut}.
The resulting $z_{\text{phot}}$ vs $z_{\text{spec}}$ distribution and associated sample statistics are presented in Fig.~\ref{fig:hzq}.

We find that our photo-$z$ estimates perform extremely well out to the highest redshifts (even extending to $z\sim7$), with the measured robust scatter $\sigma_{\text{NMAD}} = 0.05$ and outlier fraction of  $\text{OLF}_{0.15} = 2.50\%$.
Although a small number of sources have $z_{\text{phot}}$ estimates significantly below the true redshift (see e.g. the $z_{\text{spec}}\sim6.5$ source with an estimate of $z_{\text{phot}}\sim 4$), the accurate photo-$z$ errors produced by our methodology means that such catastrophic failures can be identified and excluded from scientific samples.

Given that the majority of current spectroscopically confirmed QSOs at $z > 5$ used for both training and testing were selected using traditional Lyman-break colour selection techniques, we cannot reliably quantify how our photo-$z$ estimates will perform outside of this traditional parameter space.
The fact that the photo-$z$ remain reliable at $z\sim6$, where QSOs are typically selected based on $i-z$ colours not available in the LS DR8 catalogues, suggests that the inclusion of WISE magnitude and colour information may enable such extensions.
However, spectroscopic confirmation of sources selected in this way are required to verify the full potential of the full photo-$z$ sample.
 Nevertheless, these results demonstrate that our photo-$z$ estimates can potentially be used to select robust samples of QSOs at $z > 4$ across the full $\sim19\,400$ deg$^{2}$ of LS DR8, or aid in the prioritisation or validation of candidates selected through other criteria.
 
\subsubsection{X-ray sources}
With the launch of the extended ROentgen Survey with an Imaging Telescope Array (eROSITA) on the Spectrum-Roentgen-Gamma telescope \citep{2021A&A...647A...1P}, we are now in a new era of sensitive all-sky X-rays.
In the coming years, spectroscopic surveys targeting X-ray point sources and X-ray selected cluster members \citep{2019Msngr.175...42M,2019Msngr.175...39F} will yield highly complete spec-$z$ samples discovered in the eROSITA All-Sky Survey (eRASS).
However, precise and unbiased photo-$z$s still remain a key tool for exploiting eRASS across a wide range of scientific objectives.
With LS imaging offering the deepest available all-sky (or full extragalactic sky) optical photometry available for eRASS source identification \citep{2021arXiv210614520S}, the photo-$z$s produced in this analysis offer a potentially valuable resource.
We therefore investigate how our photo-$z$ estimates perform for X-ray selected population present in the LS DR8 catalogues.

For bright X-ray sources over a large area, we consider sources from the Second Rosat all-sky survey \citep[2RXS;][]{2016A&A...588A.103B} and the XMM-Newton slew survey (XMMSL2)\footnote{\hyperlink{https://www.cosmos.esa.int/web/xmm-newton/xmmsl2-ug}{https://www.cosmos.esa.int/web/xmm-newton/xmmsl2-ug}}.
Robust optical host-identifications and spec-$z$s for the subset of these with optical spectroscopy are taken from the SPectroscopic IDentification of eROSITA Sources \citep[SPIDERS;][]{2017MNRAS.469.1065D} survey, completed as part of SDSS-IV.
To supplement the bright X-ray catalogues, we also include X-ray sources from the deep XMM-XXL North \citep{2016MNRAS.457..110M,2016MNRAS.459.1602L} and Chandra Deep Wide Field Survey \citep[CDWFS;][]{2020ApJS..251....2M} fields.
In total, the test sample of X-ray soruces consists of 16,004 sources spanning a broad range of X-ray flux densities, including significant statistics down to X-ray fluxes as faint as $F_{0.5-10~\text{keV}} \sim 3\times10^{-15}$ erg\,cm$^{-2}$\,s$^{-1}$, below the expected final eROSITA survey depths \citep[eRASS:8;][]{2021A&A...647A...1P}.
Of this full sample, 14\,207 have LS DR8 photometry free from artefacts or significant source blending.

\begin{figure}
\centering
 \includegraphics[width=0.95\columnwidth]{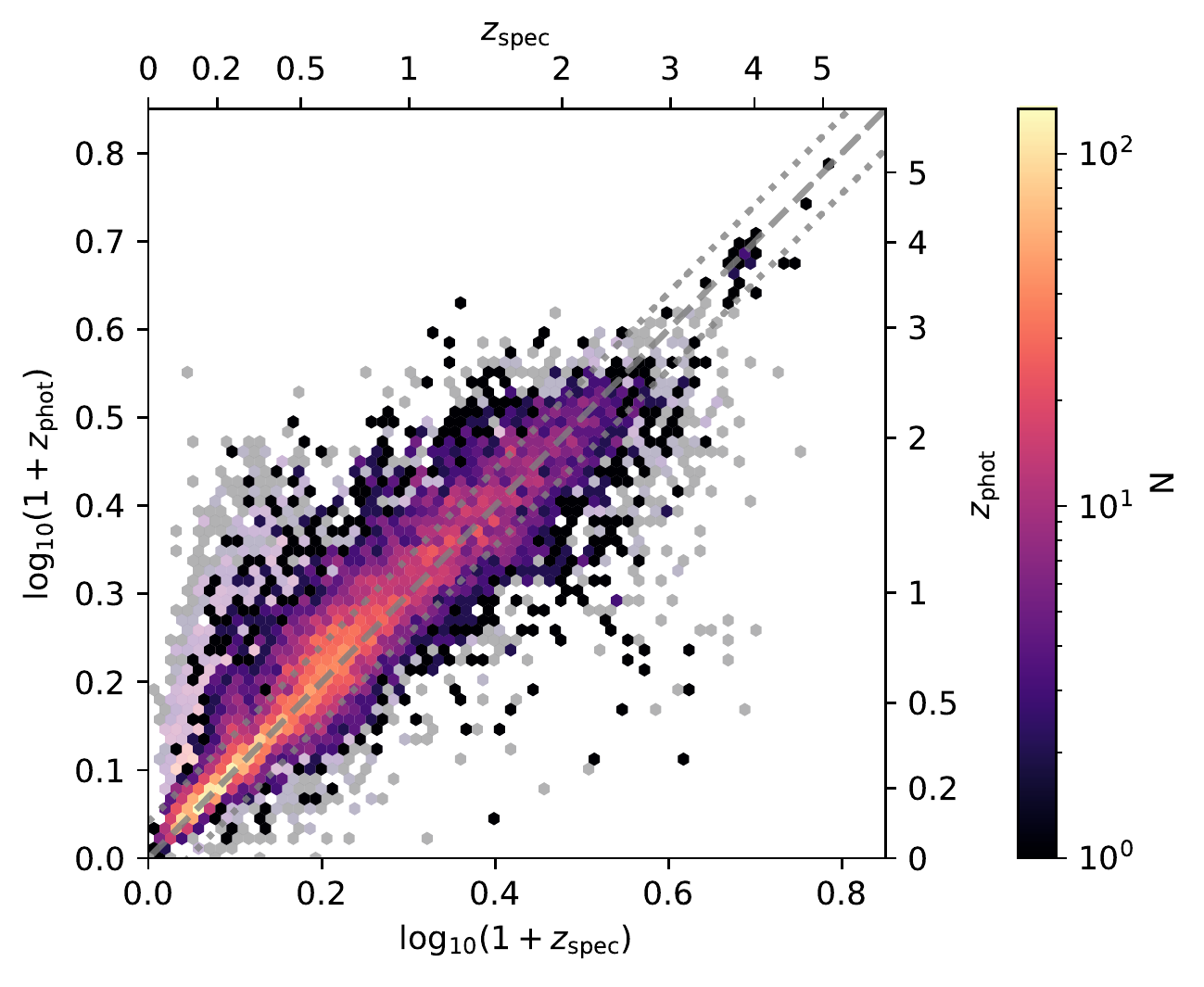}
 \caption{Distribution of predicted photo-$z$ ($z_{\text{phot}}$) vs spec-$z$ ($z_{\text{spec}}$) for a sample of X-ray detected sources from a compilation of deep X-ray surveys within the LS DR8 footprint. The colour scale illustrates the density of sources (with logarithmic scaling), with all sources with clean photometry shown as semi-transparent cells in the background and photo-$z$ estimates satisfying Eq.~\ref{eq:err_cut} over-plotted as solid colours.}
 \label{fig:xr_spz_phz}
\end{figure}

Overall, we find that the fraction of sources for which we can derive well-constrained photo-$z$ estimates is significantly lower than for the general population, with only 55\% of sources with clean photometry having $\sigma_{\text{phot}}/(1+z_{\text{phot}}) < 0.2$.
However, when examining the demographics of the X-ray spectroscopic sample, we find that a significant fraction of the sample are PSF sources at $z_{\text{spec}} < 0.5$ where our photo-$z$ estimates are known to be less reliable (Section~\ref{sec:stats_overall}).
The increased uncertainty at $z < 0.5$ can be seen in the distribution of $z_{\text{phot}}$ estimates as a function of $z_{\text{spec}}$ presented in Fig.~\ref{fig:xr_spz_phz}, where all sources in the X-ray sample with clean photometry are shown in the background, while sources with well-constrained photo-$z$ estimates are over-plotted in solid colour.
Crucially, we find that by imposing cuts based solely on the predicted photo-$z$ uncertainties (Eq.~\ref{eq:err_cut}), we can reliably select samples with accurate and unbiased photo-$z$ estimates.
In this regard, the performance for luminous X-ray sources is comparable to that of X-ray faint sources with equivalent optical properties.

\begin{figure}
\centering
 \includegraphics[width=0.95\columnwidth]{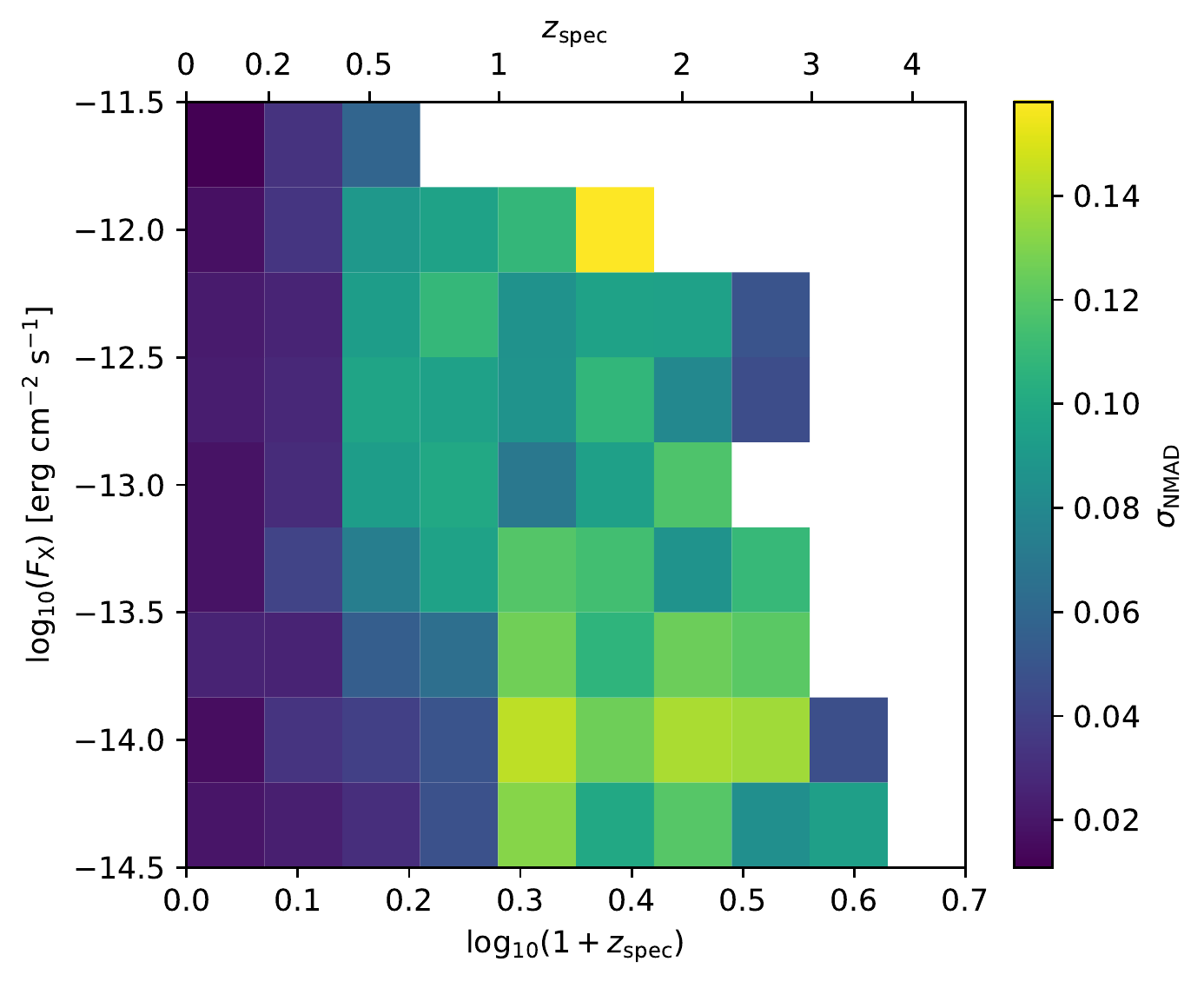}
 \includegraphics[width=0.95\columnwidth]{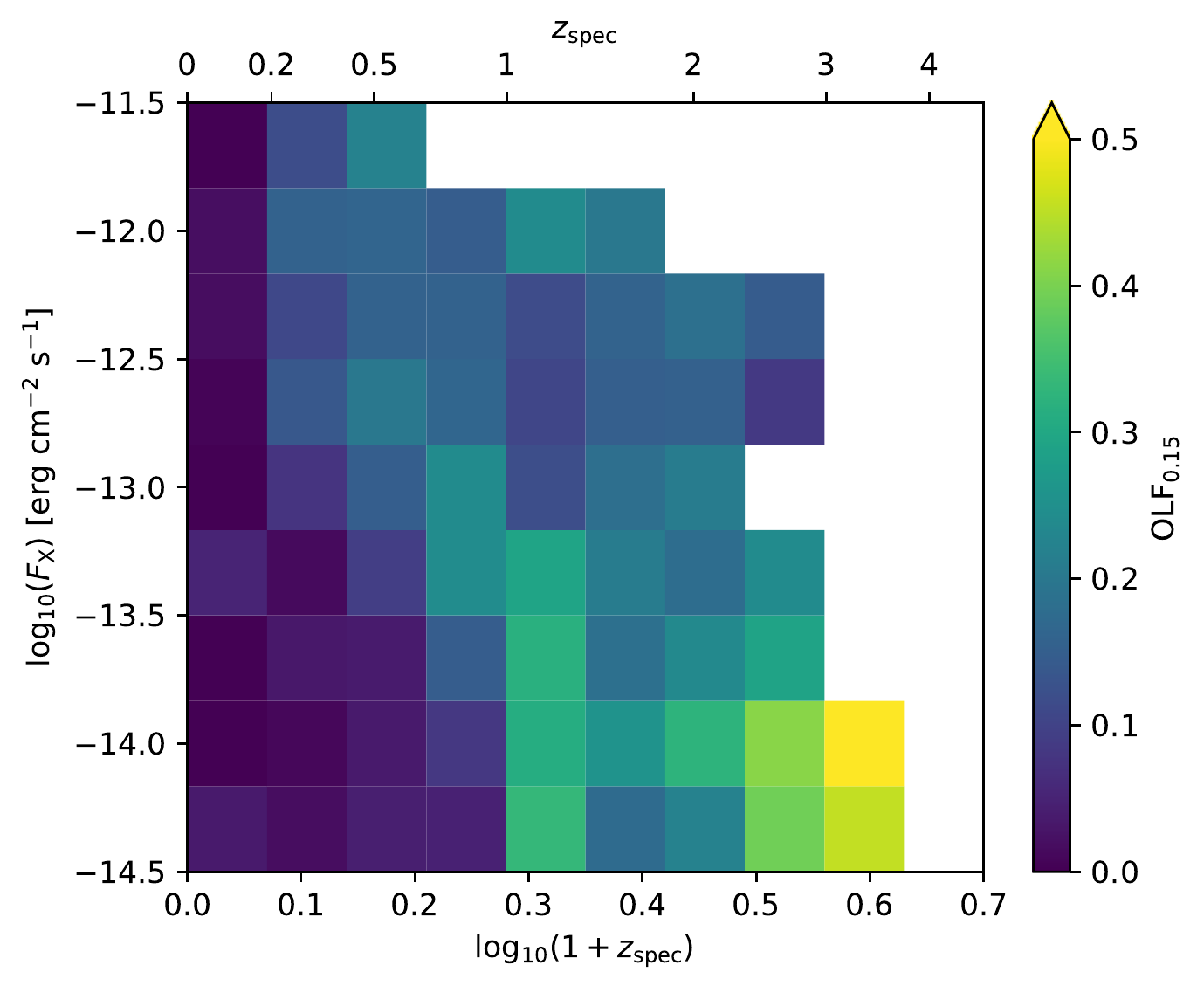}
 \caption{Robust scatter, $\sigma_{\text{NMAD}}$ (top), and absolute outlier fraction, $\text{OLF}_{0.15}$ (bottom), as a function of X-ray flux, $F_{X}$ ($ = F_{0.5-10~\text{keV}}$ or $F_{0.5-10~\text{keV}}$), and true redshift for the compilation of X-ray detected sources. Statistics are calculated only for sources with $z_{\text{phot}}$ estimates deemed as well-constrained, with a minimum of ten sources required per bin.}
 \label{fig:xr_stats_z_flux}
\end{figure}

To further investigate the dependence of our photo-$z$ quality on X-ray properties, we calculate the photo-$z$ quality as a function of observed X-ray flux, $F_{X}$ (specifically, $F_{0.5-10~\text{keV}}$ or $F_{0.5-7~\text{keV}}$ for CDWFS), and spectroscopic redshift.
The resulting statistics are presented in Fig.~\ref{fig:xr_stats_z_flux}, with the sources spanning 3 orders of magnitude in X-ray flux and extending to $z > 3$.
At $z < 0.5$, we find that the photo-$z$ predictions with low uncertainties exhibit low scatter ($\sigma_{\text{NMAD}} < 0.05$) and low outlier fraction ($<10\%$) across a very wide dynamic range in X-ray flux densities.
As redshift increases, we find that both scatter and outlier fraction increase in line with trends observed for the wider optically selected population.
Overall, our photo-$z$ estimates perform comparably to dedicated template based estimated derived using the same photometric data \citep{2021arXiv210614520S}, with the key additional benefit of also providing high quality estimates for the full optical population (Section~\ref{sec:stats_overall}).
Our photo-$z$ catalogues therefore offer the ability to e.g. robustly separate luminous background X-ray point-sources from foreground cluster members \citep{2021arXiv210614519K,2021arXiv211009544B} within one single homogeneous resource across the full sky.

\subsubsection{Radio continuum detected sources}
As with eROSITA and the X-ray sky, a new generation of wide area radio continuum surveys with facilities such as LOFAR and ASKAP are probing an order of magnitude fainter than previous all-sky surveys \citep[e.g.][]{Shimwell:2017ch}.
Together, the LOFAR Two-metre Sky Survey \citep[LoTSS;][]{Shimwell:2018to} in the northern hemisphere and the Evolutionary Map of the Universe \citep[EMU;][]{Norris:2011de} in the south will detect tens of millions of faint radio continuum sources.
The extreme diversity in optical properties for radio continuum selected sources means that these new samples will range from low-luminosity radio AGN and star-forming galaxies in the nearby Universe through to the most distant luminous quasars at high redshift \citep{2021arXiv211006222G}.

To measure the quality of the photo-$z$ estimates as a function of their radio properties we construct a sample of radio continuum selected sources with spectroscopic redshifts from the LOFAR Two-metre Sky Survey First Data Release \citep[LoTSS DR1][]{Shimwell:2018to,2019A&A...622A...2W} over $>400\,\text{deg}^{2}$.
The wide area sample from LoTSS DR1 is then supplemented with additional faint radio sources from the Bo\"{o}tes field of the LoTSS Deep Fields First Data Release \citep[][]{Tasse:2020tr,Sabater:2020ur,Kondapally2020,Duncan:21}.
The combined sample contains 32\,495 sources at $0.01 < z < 5$, of which 26\,957 have LS DR8 photometry free from artefacts and significant blending and the radio continuum flux densities range from $S_{\nu, 144\text{MHz}} <80\mu$Jy to $>10$\,Jy  (>5$\sigma$ detections) - spanning the full dynamic range of the radio continuum population.

\begin{figure}
\centering
 \includegraphics[width=0.95\columnwidth]{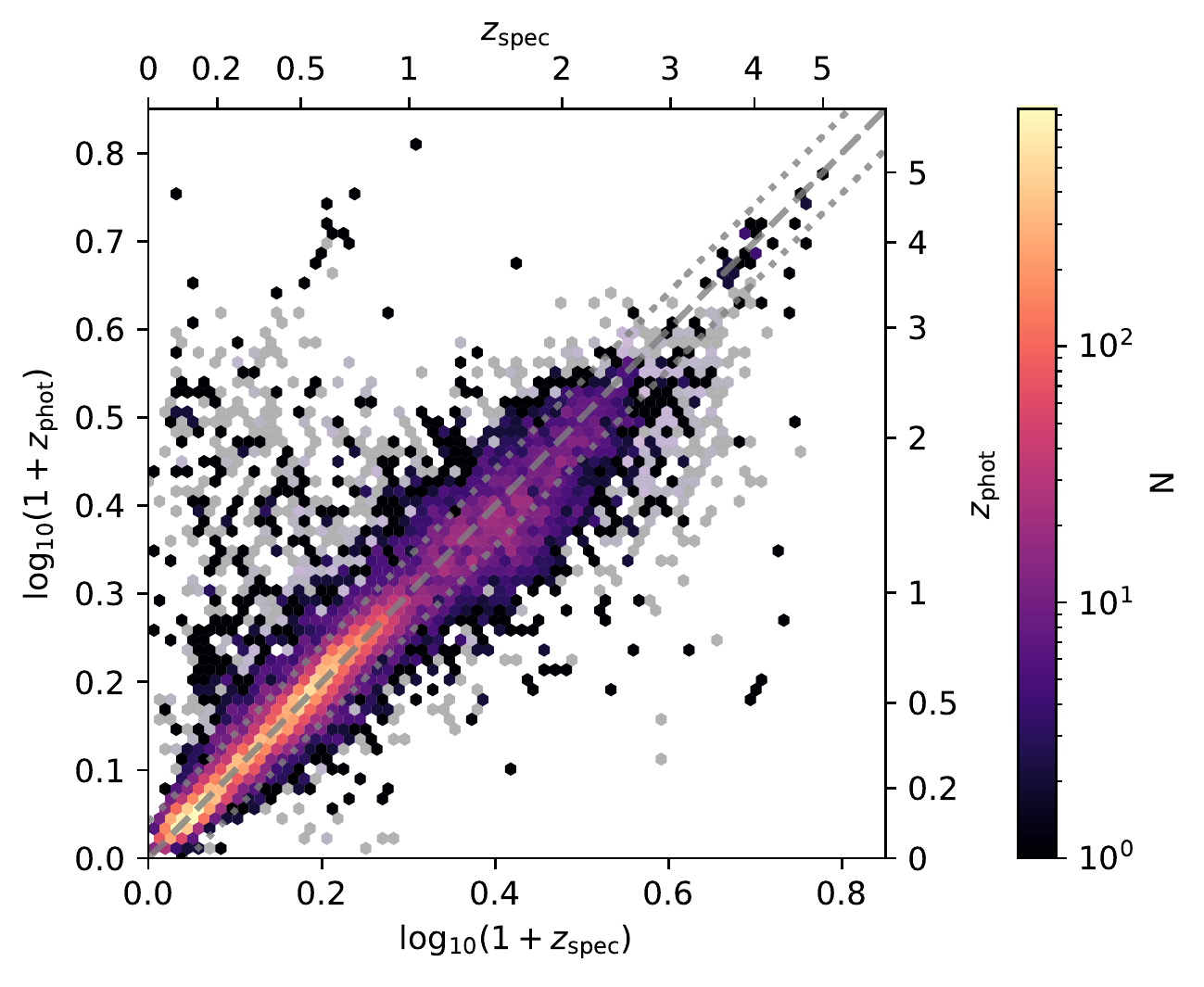}
 \caption{Distribution of predicted photo-$z$ ($z_{\text{phot}}$) vs spec-$z$ ($z_{\text{spec}}$) for a sample of low-frequency radio continuum detected sources within the LS DR8 North catalogues. The colour scale illustrates the density of sources (with logarithmic scaling), with all sources with clean photometry shown as semi-transparent cells in the background and well-constrained photo-$z$ estimates over-plotted as solid colours.}
 \label{fig:rad_spz_phz}
\end{figure}

In Fig.~\ref{fig:rad_spz_phz} we show the distribution of photo-$z$ vs spec-$z$ distribution for the radio continuum sample.
As above, we show all radio sources sample with clean photometry shown in the semi-transparent background, while sources photo-$z$ uncertainties below our permissive cut of $\sigma_{\text{phot}}/(1+z_{\text{phot}}) < 0.2$ are plotted in solid colour. 
Over the full spectroscopic sample, we find that 89\% of the radio spec-$z$ sample meet this additional uncertainty criteria, significantly more than for the X-ray selected population and indicative of the fact that the radio spec-$z$ population is dominated by resolved sources at $z < 1$ for which our photo-$z$ estimates are highly reliable. 
Similar to the X-ray population, we find that the additional photo-$z$ precision criteria excludes a significant number of $z_{\text{spec}} < 0.5$ sources with photo-$z$ estimates $z_{\text{phot}} > 1$.

\begin{figure}
\centering
 \includegraphics[width=0.95\columnwidth]{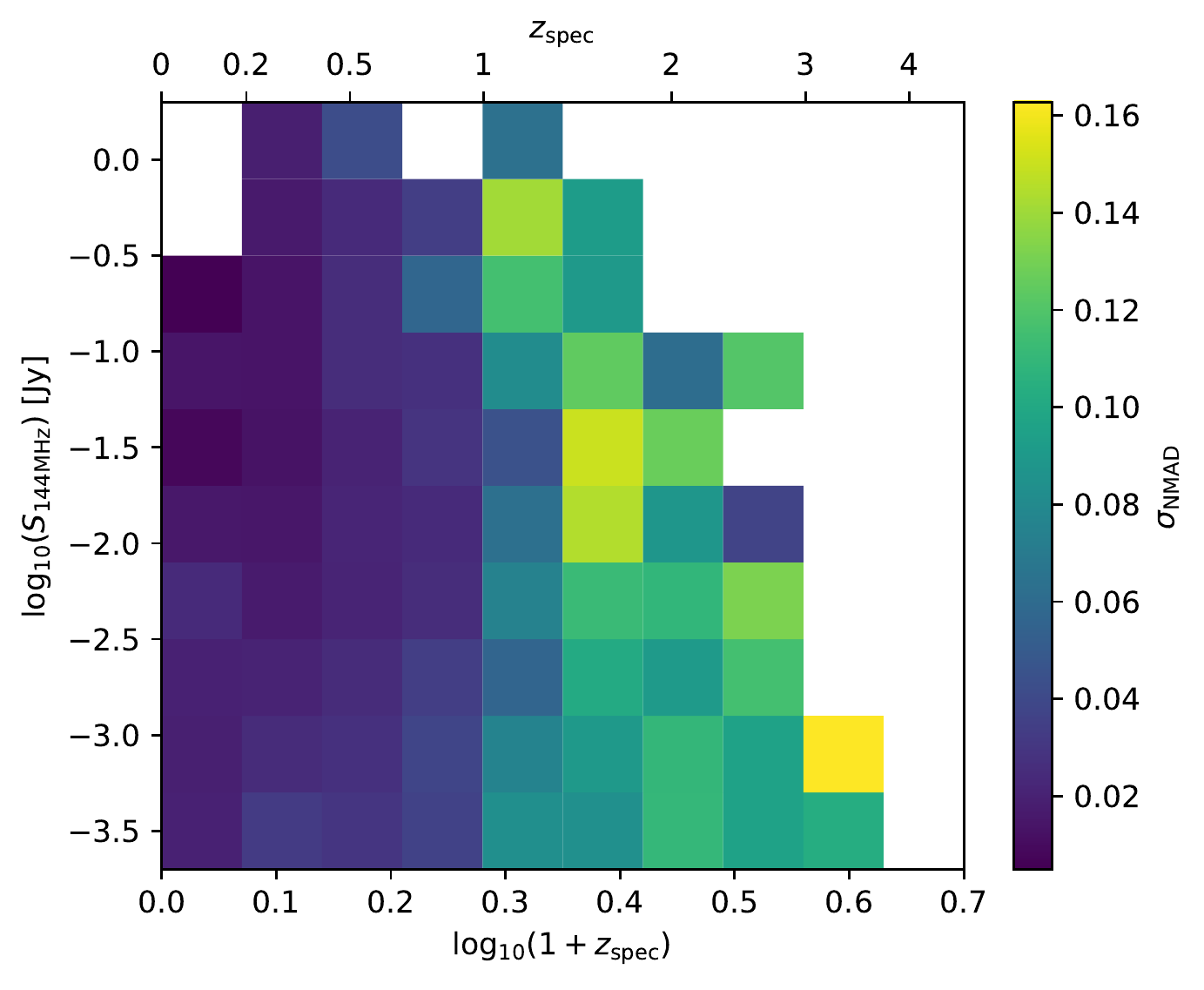}
 \includegraphics[width=0.95\columnwidth]{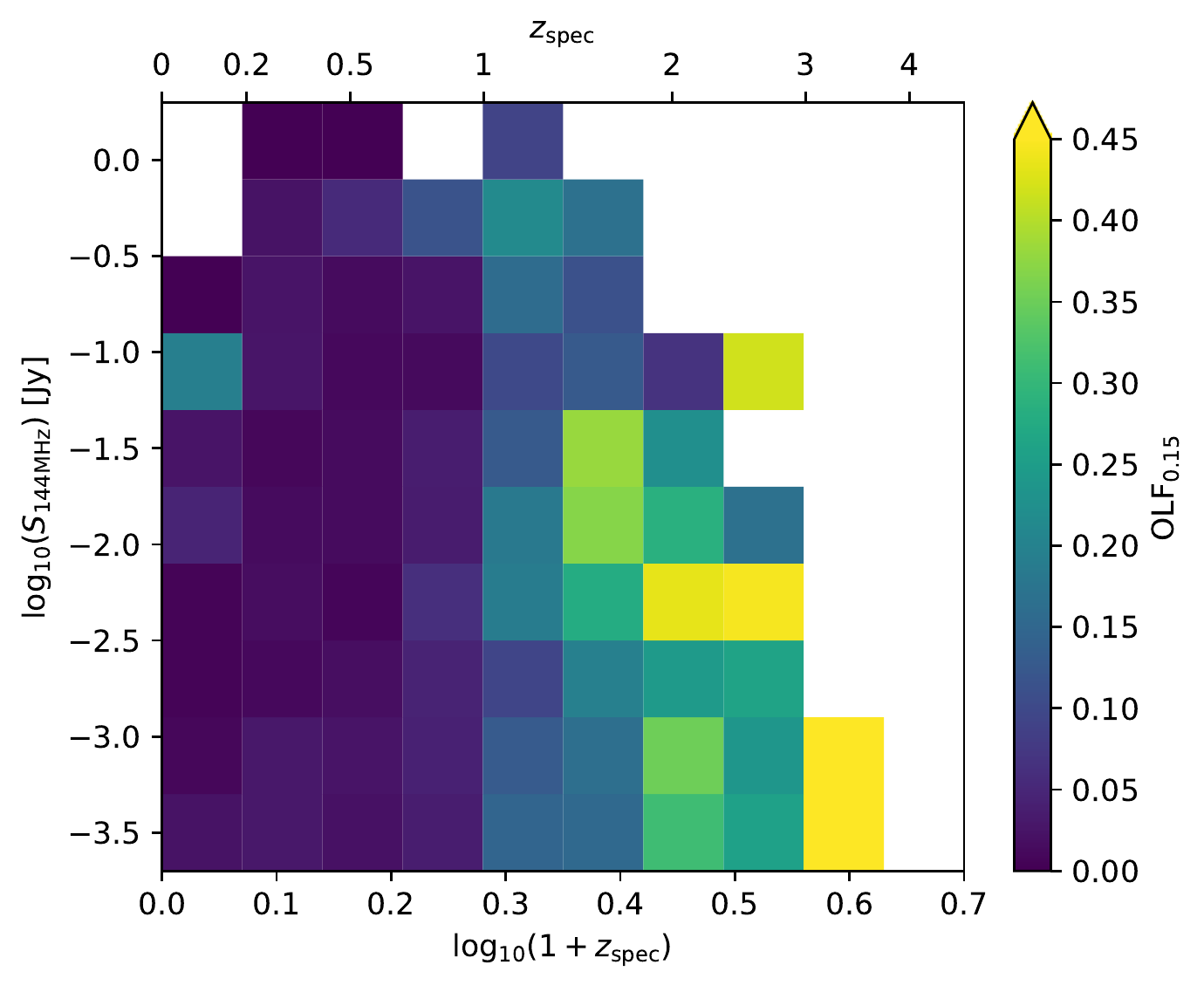}
 \caption{Robust scatter, $\sigma_{\text{NMAD}}$ (top), and absolute outlier fraction, $\text{OLF}_{0.15}$ (bottom), as a function of radio continuum flux density and true redshift for the LoTSS 144 MHz radio continuum detected sources with available spec-$z$. Statistics are calculated only for sources with photo-$z$ estimates deemed as well-constrained, with a minimum of ten sources required per bin.}
 \label{fig:rad_stats_z_flux}
\end{figure}

Fig~\ref{fig:rad_stats_z_flux} presents the $\sigma_{\textup{NMAD}}$ and $\text{OLF}_{0.15}$ as a function of both spec-$z$ and 144\,MHz flux density, $S_{\nu, 144\text{MHz}}$. 
Similar to trends observed by \citet{2019A&A...622A...3D}, we find a clear evolution in the scatter and OLF of the radio source population with $z_{\textup{spec}}$, with photo-$z$ estimates of high redshift sources being significantly worse than for sources with similar radio flux density at low redshift.
However, for a given $z_{\textup{spec}}$ bin we find no evidence of strong variation in photo-$z$ precision or reliability as a function of radio continuum flux density (and hence luminosity).

Compared directly to the photo-$z$ estimates provided by \citet{2019A&A...622A...3D} for the same sample, we find that our photo-$z$s for the optically luminous QSO population perform marginally worse.
For sources in LoTSS DR1 estimates at $1 < z_{\text{spec}} < 3$ with reliable photo-$z$ in this work, the estimates from \citet{2019A&A...622A...3D} have $\sigma_{\textup{NMAD}} = 0.091$ and $\text{OLF}_{0.15} = 19.5\%$, while the photo-$z$s in this work have $\sigma_{\textup{NMAD}} = 0.103$ and $\text{OLF}_{0.15} = 22.8\%$.
We postulate that the reasons for \citet{2019A&A...622A...3D} performing better in this regime are two-fold: firstly, the inclusion of additional $i$ and $y$-band photometry in the \textsc{GPz} estimates, and secondly, due to training \textsc{GPz} estimates specifically for identified optical QSO population (cf. the more generalised approach in this work).
Outside of this specific parameter space, where the overall reduction in quality is only of order $\sim10\%$, the photo-$z$s produced in this work significantly out-perform those provided by \citet{2019A&A...622A...3D}.
The estimates in this work have $\sim35\%$ lower scatter and outlier fraction at $z_{\text{spec}} <1$ and $\sim3-4$ times lower scatter and bias at $z_{\text{spec}} > 3$.
Given the overall performance of our photo-$z$s for radio continuum selected sources, the catalogues produced in this work offer a valuable resource for exploiting the new generation of wide area radio continuum surveys.

\section{Photo-$z$ predictions for the full LS DR8}\label{sec:catalogues}

Following the validation of the final photo-$z$ performance and reliability, photo-$z$ predictions for the full LS DR8 sample were produced by processing each of the 123 and 349 distinct \textsc{HEALPix} regions for LS DR8 North and South respectively with the pipeline steps outlined in Section~\ref{sec:final-method}.
As outlined in Section~\ref{sec:ls_data}, catalogues were limited to LS DR8 bricks with at least one exposure in all three optical bands and luminous bright \textit{Gaia} sources (\texttt{type} = `DUP') were excluded.
In total, 323\,213\,867 and 1\,252\,523\,992 sources LS DR8 North and South respectively were processed through the pipeline.
However, these full optical photometry catalogues will contain many sources for which photo-$z$ estimates are not valid, or reliable.
This may be the case either because the source is a star, or because the photometric is adversely effected by imaging artefacts, neighbouring bright stars (e.g. through diffraction spikes) or due to very significant blending between close objects.

To allow for high quality scientific samples to be easily selected from the full photo-$z$ catalogues, we define a set of simple criteria to select sources for which photo-$z$ estimates are expected to be relatively unbiased and accurate enough for general astrophysical studies.
Photometry for a given object is deemed to be `clean' if the following criteria are satisfied:
\begin{equation}\label{eq:blend}
\texttt{fracflux\_}x < 0.33 \quad \text{or} \quad S/N(x) < 2 
\end{equation}
and 
\begin{equation}\label{eq:mask}
\texttt{maskbits} = 0
\end{equation}
where $x$ is each of the three optical bands, $g$, $r$ and $z$.
Allowing $\texttt{fracflux\_}x > 0.33$ for objects which have $S/N(x) < 2$ ensures that objects which are genuinely non-detections in a particular band (e.g. high-redshift sources) are not excluded due to low levels of contaminating flux from neighbouring objects.
In addition, to exclude likely stars we require 
\begin{equation}\label{eq:pstar_cut}
P_{\text{star}} < 0.2,	
\end{equation}
for unresolved optical sources. 
Finally, estimates are deemed plausible if they satisfy the uncertainty criteria outlined in Eq.~\ref{eq:err_cut}.

One of the key advantages of the \textsc{GPz} algorithm is its ability to account for noise from the training data when deriving uncertainty estimates. 
Sources with properties not well modelled by the spectroscopic training sample (i.e. stars) therefore have large predicted uncertainties, even if the noise on the input magnitudes are extremely small.
Examining the \textsc{GPz} predictions for spectroscopically confirmed stars in LS DR8 (from SDSS DR14, Section~\ref{sec:star-gal}), we find that using solely the photo-$z$ uncertainty cut is able to successfully remove 93.9\% of SDSS stars from the sample.
Requiring $P_{\text{star}} < 0.2$ alone successfully excludes 98.9\% of stars.
However, when combined the criteria in Eq.~\ref{eq:pstar_cut} and \ref{eq:err_cut} are able to exclude 99.9\% of stars in the SDSS spectroscopic sample. 
Exploring the effectiveness of these cuts as a function of optical magnitude we find that the purity does decline for fainter sources, ranging from 100\% of SDSS stars excluded at bright magnitudes down to $97.0^{+1.3}_{-2.2}\,\%$ at $m_{z} = 21.5$ (where the uncertainty is derived solely from binomial uncertainties from sample statistics).

\begin{table}
\centering
\caption{Total sample sizes for the full optical catalogues processed in this analysis, the subset which pass the criteria for `clean' photometry (Eq.~\ref{eq:blend} and \ref{eq:mask}), and the number predicted to have good quality photo-$z$ estimates based on the combination of quality criteria outlined in the text.}
\begin{tabular}{lrrr}
\hline
Morphology & Total & Clean Photometry &  `Good' Photo-$z$\\
\hline
& \multicolumn{3}{c}{North} \\
\hline
PSF & 135\,630\,162 & 109\,369\,165 & 41\,938\,327 \\
DEV & 15\,445\,582 & 13\,058\,724 &13\,050\,326 \\
EXP & 24\,354\,533 & 21\,777\,404 & 21\,388\,924 \\
REX & 147\,576\,416 & 132\,511\,841 & 123\,654\,049 \\
COMP & 207\,174 & 145\,098 & 127\,708 \\
\hline
All & 323\,213\,867 & 276\,862\,232 & 200\,159\,334 \\
\hline
& \multicolumn{3}{c}{South} \\
\hline
PSF & 571\,512\,918 & 489\,858\,010 & 205\,831\,876 \\
DEV & 42\,818\,253 & 37\,032\,451 & 37\,006\,782 \\
EXP & 114\,955\,799 & 105\,469\,819 & 100\,327\,246 \\
REX & 522\,373\,297 & 477\,763\,969 & 406\,110\,027 \\
COMP & 863\,725 & 660\,308 & 595\,708 \\
\hline
All & 1\,252\,523\,992 & 1\,110\,784\,557 & 749\,871\,639 \\
\hline
\end{tabular}\label{tab:photz_summary}
\end{table}

Table~\ref{tab:photz_summary} summarises the total photometric sample processed in this analysis for both LS DR8 North and South, as well as the number of sources that satisfy the quality criteria outlined above - deemed to be `good' photo-$z$ predictions.
Combining the two samples following the approach taken by the LS DR8 team (where LS DR8 North sources are only included at $\delta > 32.375\degr$ and in the North galactic plane), the total number of unique sources with `good' photo-$z$ estimates is 936\,131\,489.
Averaged over all source morphologies, the fraction of sources with good photo-$z$ estimates is similar between the two datasets (61.9\% and 59.9\% in the North and South respectively).
As expected, significantly fewer sources with PSF morphology have good photo-$z$ predictions, consistent with the majority of bright point-sources in the LS DR8 optical catalogues being stars.
The fraction of PSF sources with good photo-$z$s is also seen to vary strongly with galactic latitude, ranging from only $\sim15\%$ of PSF sources with good estimates at galactic latitudes of $| b | < 20\degr$, up to $\approx 45\%$ at $| b | > 80\degr$.
Similar trends with galactic latitude are seen for resolved morphologies, but with weaker evolution.
For example, the fraction of exponential morphologies with good photo-$z$ estimates ranges from $\approx 70\%$ to $\approx 90\%$ over the same range.
The decline in photo-$z$ reliability for resolved sources is driven primarily by the impacts of increasing source density (particularly of bright sources), with half of the increase in poor photo-$z$s attributed to the increased fraction of sources affected by masked pixels (e.g. due to bright stars, diffraction spikes etc.).

\begin{figure}
\centering
 \includegraphics[width=0.99\columnwidth]{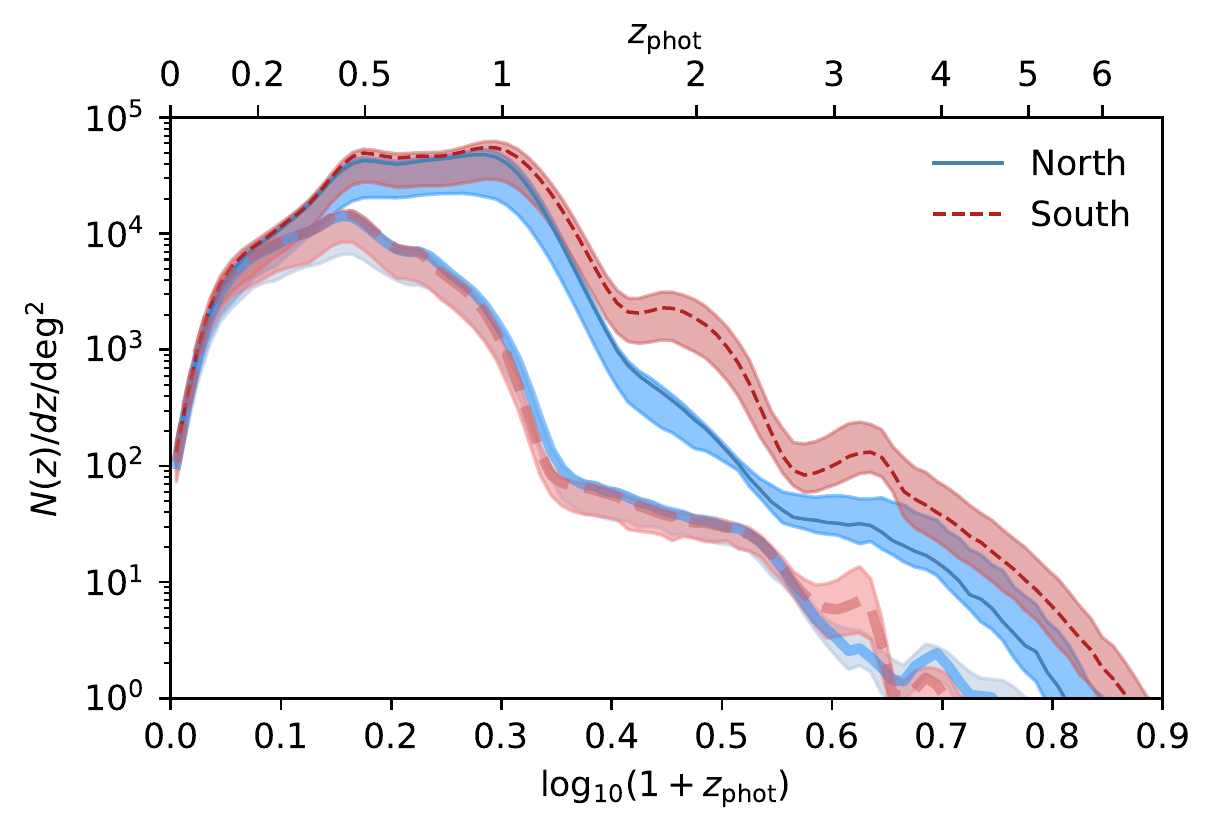}
 \caption{Distribution of photo-$z$ point-estimates for sources in the LS DR8 catalogues with good quality photo-$z$ estimates. Solid and dashed lines represent the median source density per unit redshift across all \textsc{HEALPix} regions in LS DR8 North and South respectively, with shaded regions illustrate the corresponding inter-quartile ranges. Thick lines show the median source density per unit redshift for sources with $m_{z} < 21$, illustrating that the two datasets produce consistent predictions for the same magnitude limit.}
 \label{fig:photoz_dist}
\end{figure}

When comparing statistics for the two datasets, the most significant difference occurs for round exponential morphologies (REX), which constitute the majority of resolved sources in both datasets. 
Given the excellent consistency between North and South for both spec-$z$ test samples and with each other (see Appendix~\ref{app:pz_offsets}), we attribute the lower fraction of good estimates in LS DR8 South to the increased optical depth resulting in the detection greater numbers of higher redshift ($z > 1$) sources, for which the photo-$z$ estimates are less precise.
In Fig.~\ref{fig:photoz_dist} we show the distribution of $z_{\text{phot}}$ per unit area of sources with good photo-$z$ estimates in both datasets (not accounting for the full posterior predictions of individual sources).
At $z < 1$, where resolved morphologies constitute a large fraction of the extragalactic population, the photo-$z$ distributions (averaged over all \textsc{HEALPix} regions) for North and South are in extremely strong agreement.
Above $z > 1$, LS DR8 South provides additional well-constrained photo-$z$ estimates compared to North, but follows a similar decline in the source density of robust estimates as redshift increases -- in both datasets the good photo-$z$ estimates extend beyond $z > 6$ (see Section~\ref{sec:results_highz}).
The extremely tight agreement in the predicted redshift distribution of bright optical samples ($m_{z} < 21$; thick solid and dashed lines respectively) supports the conclusion that the primary cause of the difference in predictions for LS DR8 North and South is the depths of the optical photometry.

\subsection{Photo-$z$ catalogues}
Table~\ref{tab:cat_desc} presents information on the columns available in the photo-$z$ catalogues, while Table~\ref{tab:cat_example} provides an example of the catalogues for the first ten sources in LS DR8.
Key object identifier and positional information from the underlying LS DR8 photometric catalogues is included to enable unambiguous matching to other columns available in LS DR8 (including underlying flux information and additional morphological properties).
The catalogues provided in this work contain information for all optical sources processed through the \textsc{GPz} pipeline, regardless of the reliability of the resulting estimates.
Sources that appear in both LS DR8 North and South optical catalogues will have separate photo-$z$ predictions based on the two sets of photometry and we provide all available predictions.
For consistency with the convention used by the Legacy Survey collaboration we recommend users follow the approach used above, using LS DR8 North predictions for sources at $\delta > 32.375\degr$ in the North Galactic Cap ($ b > 0 \degr$) and LS DR8 South predictions otherwise.

In addition to the main photo-$z$ prediction and calibrated 1-$\sigma$ uncertainties, a separate flag is provided for all sources indicating whether they satisfy the criteria for `good' photo-$z$ prediction (Eq.10--13). 
As outlined above, this photo-$z$ quality flag is able to reliably exclude most bright stars and extragalactic sources with unreliable estimates, allowing for robust samples over a wide range of parameter space.
However, we caution that with the full sample reaching over $\gtrsim10^{9}$ sources, no single assessment criteria can offer 100\% reliability for all potential scientific uses.
We therefore encourage users of the photo-$z$ catalogues to incorporate additional sample selection criteria appropriate for their specific target population and scientific goals.

\begin{table*}
\caption{Format of the LS DR8 North and South photo-$z$ catalogues produced in this work. Full catalogues can be accessed online through the \textit{Vizier} catalogue service as well as the NOIRLab \textit{Astro Data Lab}.}\label{tab:cat_desc} 
\begin{tabular}{lll}
	\hline
Column & Units & Description \\
\hline 
 \texttt{id} & & Unique object identifier combining \texttt{release}, \texttt{brickid} and \texttt{objid} \\
 \texttt{release} & & LS DR8 release number - denoting the camera and filter set used \\
 \texttt{brickid} & & LS DR8 brick number [1,662174] \\
 \texttt{objid} & & LS DR 8 catalogue object number within this brick\\
 \texttt{ra} & deg & Right ascension (J2000)  \\
  \texttt{dec} & deg & Declination (J2000) \\
  \texttt{type} & & Optical morphological model: PSF, REX, DEV, EXP or COMP\\
  \texttt{pstar} & & $P_{\text{star}}$ - star likelihood based on colours from GMM star-QSO classification (Valid only for PSF sources) \\
  \texttt{gmmcomp} & & Best-matching GMM component used to derive photo-$z$ estimate\\
 \texttt{zphot} & & Photo-$z$ estimate (mean of the normally distributed photo-$z$ posterior)  \\
\texttt{zphot\_err} & & Uncertainty on photo-$z$ estimate (standard deviation of the normally distributed photo-$z$ posterior) \\
\texttt{flag\_clean} & & Photometry reliability flag, $=1$ for sources free of blending or imaging artefacts (Eq.~\ref{eq:blend} and \ref{eq:mask}).\\
\texttt{flag\_qual} & & Photo-$z$ reliability flag, 1 for sources expected to have well-constrained estimates (\texttt{flag\_clean} and satisfying Eq.~\ref{eq:err_cut} and \ref{eq:pstar_cut}).\\
\hline
\end{tabular}\\
\end{table*}

\setlength\tabcolsep{3pt}

\begin{table*}
\caption{Example of the LS DR8 photo-$z$ catalogues produced in this work, showing the first 10 entries of the North catalogue. Column descriptions are as outlined in Table~\ref{tab:cat_example}.}\label{tab:cat_example} 
\footnotesize
\begin{tabular}{ccccccccccccc}
\hline
\texttt{id} & \texttt{release} & \texttt{brickid} & \texttt{objid} & \texttt{ra} & \texttt{dec} & \texttt{type} & \texttt{pstar} & \texttt{gmmcomp} & \texttt{zphot} & \texttt{zphot\_err} &  \texttt{flag\_clean} & \texttt{flag\_qual} \\
\hline
8001655327000001 & 8001 & 655327 & 1 & 134.491665 & 78.375823 & PSF & 0.056 & P3 & 1.099 & 0.202 & 1 & 1 \\
8001655327000002 & 8001 & 655327 & 2 & 134.485347 & 78.376252 & REX & 0.005 & R4 & 0.940 & 0.129 & 1 & 1 \\
8001655327000003 & 8001 & 655327 & 3 & 134.494718 & 78.376295 & REX & 0.006 & R8 & 0.949 & 0.148 & 1 & 1 \\
8001655327000004 & 8001 & 655327 & 4 & 134.533553 & 78.375083 & PSF & 0.225 & P1 & 0.808 & 0.272 & 0 & 0 \\
8001655327000006 & 8001 & 655327 & 6 & 134.517881 & 78.377070 & PSF & 0.002 & P0 & 0.676 & 0.702 & 0 & 0 \\
8001655327000007 & 8001 & 655327 & 7 & 134.516587 & 78.376066 & PSF & 0.001 & P1 & 0.715 & 0.228 & 0 & 0 \\
8001655327000008 & 8001 & 655327 & 8 & 134.488446 & 78.377107 & REX & 0.001 & R5 & 0.909 & 0.242 & 1 & 1 \\
8001655327000013 & 8001 & 655327 & 13 & 135.463593 & 78.375506 & REX & 0.003 & R0 & 0.927 & 0.252 & 1 & 1 \\
8001655327000021 & 8001 & 655327 & 21 & 134.586454 & 78.376363 & REX & 0.000 & R4 & 0.515 & 0.079 & 1 & 1 \\
8001655327000023 & 8001 & 655327 & 23 & 134.575900 & 78.377192 & DEV & 0.003 & D3 & 0.414 & 0.098 & 1 & 1 \\
\hline
\end{tabular}
\end{table*}

\subsection{LS Data Release 9 and future prospects}
The photo-$z$ catalogues produced in this work are based on DR8 of the LS optical catalogues, the latest available at the point predictions for the full optical samples commenced (Section~\ref{sec:catalogues}).
Subsequent to this analysis, Legacy Surveys DR9\footnote{\href{https://www.legacysurvey.org/dr9/}{https://www.legacysurvey.org/dr9/}} was made public. 
While LS DR9 does not include significant new optical imaging, it does include additional NEOWISER observations that increase sensitivity in WISE bands as well incorporating several improvements to the pipeline processing.
Future applications of the method to LS DR9 therefore offers the potential for additional small improvements in photo-$z$ quality and greater sample sizes.

Independent of any changes or expansion to the underlying photometric catalogues available, the commencement of a new generation of galaxy and cosmology spectroscopic surveys means that spectroscopic training samples available for photo-$z$ estimation will grow rapidly in the coming years.
Most relevant in this landscape is the DESI survey \citep{2016arXiv161100036D}, which will provide both magnitude selected samples at low redshift and colour selected samples of cosmological tracers out to higher redshift (including emission line galaxies at $z\sim1$ and QSOs at $z > 2$).
The WEAVE-LOFAR survey \citep{Smith:2016vw} will also soon provide spec-$z$ measurements for $>10^{5}$ radio continuum selected sources, providing training samples hugely complementary to those derived from optical selection.
Additionally, surveys in the southern hemisphere with the forthcoming the 4-metre Multi-Object Spectroscopic Telescope \citep[4MOST;][]{2019Msngr.175....3D} will provide large spectroscopic samples across a broad range of galaxy and AGN types, including dedicated follow-up of eROSITA selected X-ray sources will further expand the available training samples in the coming years \citep{2019Msngr.175...42M}.

There is therefore potential for significant improvements in the precision and reliability of photo-$z$ estimates for the LS datasets (and future large area optical surveys).
The methodology outlined in this work is well suited to benefit from these new training samples, with the GMM population division and CSL weighting steps able to maximise the impact of training samples in new or previously under-sampled regions of parameter space.
As training samples increase, the optical parameter space can be further divided (or better modelled) with GMMs, either with additional components or extension to higher dimensions, while still providing sufficient training and test samples for \textsc{GPz}.
Further optimisation of the photo-$z$ methodology itself is also still possible, for example through dynamic allocation of additional GP basis functions to GMM components with significantly larger training samples (reallocating from models with significant over-fitting found from the validation sample).

\section{Summary and conclusions}\label{sec:summary}
In this paper we present new photometric redshift (photo-$z$) estimates derived for the DESI Legacy Imaging Surveys Data Release 8 (LS DR8), covering almost the full extra-galactic low-extinction sky ($\sim 19\,400$\,deg$^{2}$).
By design, our photo-$z$ estimation methodology aims to produce robust photo-$z$ predictions for all populations present in the optical catalogues, ensuring that predictions for rare but scientifically valuable populations are given comparable weight to the more numerous populations that can typically dominate traditional empirical (or machine-learning) photo-$z$ training procedures.

Our method employs Gaussian mixture models (GMMs) derived from the colour, magnitude and size properties of both the training and full optical samples.
Using these purely data-driven models, we are able to divide the observed population into different regions of parameter space and to weight the training samples based on the relative density of the training and full populations.
The sparse Gaussian processes redshift code, \textsc{GPz}, is then used to derive photo-$z$ estimates for individual regions of observed parameter space, including cost-sensitive learning weights derived from the GMMs to mitigate against biases in the spectroscopic training sample.
For unresolved optical sources (PSF morphologies in the LS DR8 catalogue), separate GMMs are used to calculate tentative star-QSO classifications based on literature spectroscopic training samples - ensuring that the primary GMM based population division and weights are not biased by contamination from stars in the optical sample.
Subsequent to photo-$z$ prediction, we perform an additional calibration step on the predicted uncertainties, significantly improving the population averaged accuracy of the uncertainty estimates.

In comparison to other photo-$z$ predictions available in the literature for the same optical population, we find that our photo-$z$ estimates offer substantially improved reliability and precision at $z > 1$, with negligible loss in accuracy for brighter, resolved populations at $z < 1$.
Examining the photo-$z$ predictions for key sub-populations of the full optical sample, we find that our photo-$z$s are able to provide extremely high quality predictions for some of the rarest populations, including QSOs out to $z > 6$ and luminous star-forming galaxies and AGN selected via deep radio continuum or X-ray observations.
With appropriate quality cuts, our photo-$z$ predictions for X-ray and radio continuum selected populations can be used over a wide range in parameter space - with low robust scatter ($\sigma_{\text{NMAD}} < 0.02 - 0.10$) and outlier fraction ($\textup{OLF}_{0.15} <10\%$) at $z < 1$ across a broad range of X-ray or radio continuum flux (densities).
The photo-$z$ catalogues provided in this work therefore offer enormous potential value for new generations of all-sky X-ray (eROSITA) and radio continuum surveys (LOFAR, ASKAP), whilst still offering extremely high precision and reliability for broader optically selected samples.

Alongside the photo-$z$ prediction and associated uncertainty for all $1.6\times10^{9}$ optical sources processed in this analysis, we provide a simple photo-$z$ reliability flag that incorporates photometric quality flags, source contamination, star-QSO probability and photo-$z$ precision. 
Combining both LS DR8 North and South datasets, the catalogues provided in this work offer high quality photo-$z$ predictions for $>9\times10^{8}$ galaxies across the bulk of the extragalactic sky, making it one of the most extensive samples of redshift estimates ever produced.

\section*{Acknowledgements}
The author thanks the referee for their feedback and contributions to improving the manuscript, Philip Best for valuable feedback on various stages of the work and Peter Hatfield and Matt Jarvis for helpful discussions early in the development of the methodology.
The author acknowledges funding from the European Union's Horizon 2020 research and innovation programme under the Marie Sk\l{}odowska-Curie grant agreement No. 892117 (HIZRAD).

The Legacy Surveys consist of three individual and complementary projects: the Dark Energy Camera Legacy Survey (DECaLS; Proposal ID \#2014B-0404; PIs: David Schlegel and Arjun Dey), the Beijing-Arizona Sky Survey (BASS; NOAO Prop. ID \#2015A-0801; PIs: Zhou Xu and Xiaohui Fan), and the Mayall z-band Legacy Survey (MzLS; Prop. ID \#2016A-0453; PI: Arjun Dey). DECaLS, BASS and MzLS together include data obtained, respectively, at the Blanco telescope, Cerro Tololo Inter-American Observatory, NSF's NOIRLab; the Bok telescope, Steward Observatory, University of Arizona; and the Mayall telescope, Kitt Peak National Observatory, NOIRLab. The Legacy Surveys project is honored to be permitted to conduct astronomical research on Iolkam Du'ag (Kitt Peak), a mountain with particular significance to the Tohono O'odham Nation.

NOIRLab is operated by the Association of Universities for Research in Astronomy (AURA) under a cooperative agreement with the National Science Foundation.

This project used data obtained with the Dark Energy Camera (DECam), which was constructed by the Dark Energy Survey (DES) collaboration. Funding for the DES Projects has been provided by the U.S. Department of Energy, the U.S. National Science Foundation, the Ministry of Science and Education of Spain, the Science and Technology Facilities Council of the United Kingdom, the Higher Education Funding Council for England, the National Center for Supercomputing Applications at the University of Illinois at Urbana-Champaign, the Kavli Institute of Cosmological Physics at the University of Chicago, Center for Cosmology and Astro-Particle Physics at the Ohio State University, the Mitchell Institute for Fundamental Physics and Astronomy at Texas A\&M University, Financiadora de Estudos e Projetos, Fundacao Carlos Chagas Filho de Amparo, Financiadora de Estudos e Projetos, Fundacao Carlos Chagas Filho de Amparo a Pesquisa do Estado do Rio de Janeiro, Conselho Nacional de Desenvolvimento Cientifico e Tecnologico and the Ministerio da Ciencia, Tecnologia e Inovacao, the Deutsche Forschungsgemeinschaft and the Collaborating Institutions in the Dark Energy Survey. The Collaborating Institutions are Argonne National Laboratory, the University of California at Santa Cruz, the University of Cambridge, Centro de Investigaciones Energeticas, Medioambientales y Tecnologicas-Madrid, the University of Chicago, University College London, the DES-Brazil Consortium, the University of Edinburgh, the Eidgenossische Technische Hochschule (ETH) Zurich, Fermi National Accelerator Laboratory, the University of Illinois at Urbana-Champaign, the Institut de Ciencies de l'Espai (IEEC/CSIC), the Institut de Fisica d'Altes Energies, Lawrence Berkeley National Laboratory, the Ludwig Maximilians Universitat Munchen and the associated Excellence Cluster Universe, the University of Michigan, NSF's NOIRLab, the University of Nottingham, the Ohio State University, the University of Pennsylvania, the University of Portsmouth, SLAC National Accelerator Laboratory, Stanford University, the University of Sussex, and Texas A\&M University.

BASS is a key project of the Telescope Access Program (TAP), which has been funded by the National Astronomical Observatories of China, the Chinese Academy of Sciences (the Strategic Priority Research Program "The Emergence of Cosmological Structures" Grant \# XDB09000000), and the Special Fund for Astronomy from the Ministry of Finance. The BASS is also supported by the External Cooperation Program of Chinese Academy of Sciences (Grant\# 114A11KYSB20160057), and Chinese National Natural Science Foundation (Grant \# 11433005).

The Legacy Survey team makes use of data products from the Near-Earth Object Wide-field Infrared Survey Explorer (NEOWISE), which is a project of the Jet Propulsion Laboratory/California Institute of Technology. NEOWISE is funded by the National Aeronautics and Space Administration.

The Legacy Surveys imaging of the DESI footprint is supported by the Director, Office of Science, Office of High Energy Physics of the U.S. Department of Energy under Contract No. DE-AC02-05CH1123, by the National Energy Research Scientific Computing Center, a DOE Office of Science User Facility under the same contract; and by the U.S. National Science Foundation, Division of Astronomical Sciences under Contract No. AST-0950945 to NOAO.

\section*{Data Availability}
The data underlying this article are available through the \href{https://www.legacysurvey.org/dr8/}{DESI Legacy Imaging Surveys consortium website}, as well as through the \href{https://datalab.noirlab.edu}{NOIRLab Astro Data Lab} service.
All catalogues produced in this work will be made available through \href{https://vizier.cfa.harvard.edu/viz-bin/VizieR}{the VizieR repository} and \href{https://datalab.noirlab.edu}{NOIRLab Astro Data Lab}.



\bibliographystyle{mnras}
\bibliography{bibtex_library} 




\appendix

%

\section{Comparison of LS DR8 North and South predictions}\label{app:pz_offsets}
To allow our photo-$z$ training to incorporate training samples across both LS DR8 North and South datasets, optical magnitudes from LS DR8 North used in this analysis are homogenised to match the photometric system of South based on the observed offsets within overlapping regions (see Section~\ref{sec:data_homogenisation}).
It is crucial to verify that these corrections do not result in any small-scale biases that could impact scientific analysis of samples that span both datasets.
While no evidence for statistical biases in the resulting photo-$z$ estimates can be seen in comparison with corresponding spec-$z$ estimates, the spec-$z$ test sample represents only a limited subset of the full photo-$z$ catalogues.
We therefore perform a direct comparison between the two photo-$z$ predictions derived for sources in the region with overlapping observations from both LS DR8 North and South.

For a random subset of the full sample of galaxies in the overlapping region, we positionally cross-matched the output photo-$z$ catalogues - keeping matches within a conservative limit of 0.35\arcsec (the pixel scale of the underlying imaging).
Limiting our analysis to only those sources with reliable photo-$z$ estimates in both catalogues ($\texttt{flag\_qual} =1$), we explore the difference in predicted photo-$z$ as a function of a number of properties.

\begin{figure}
\centering
 \includegraphics[width=0.98\columnwidth]{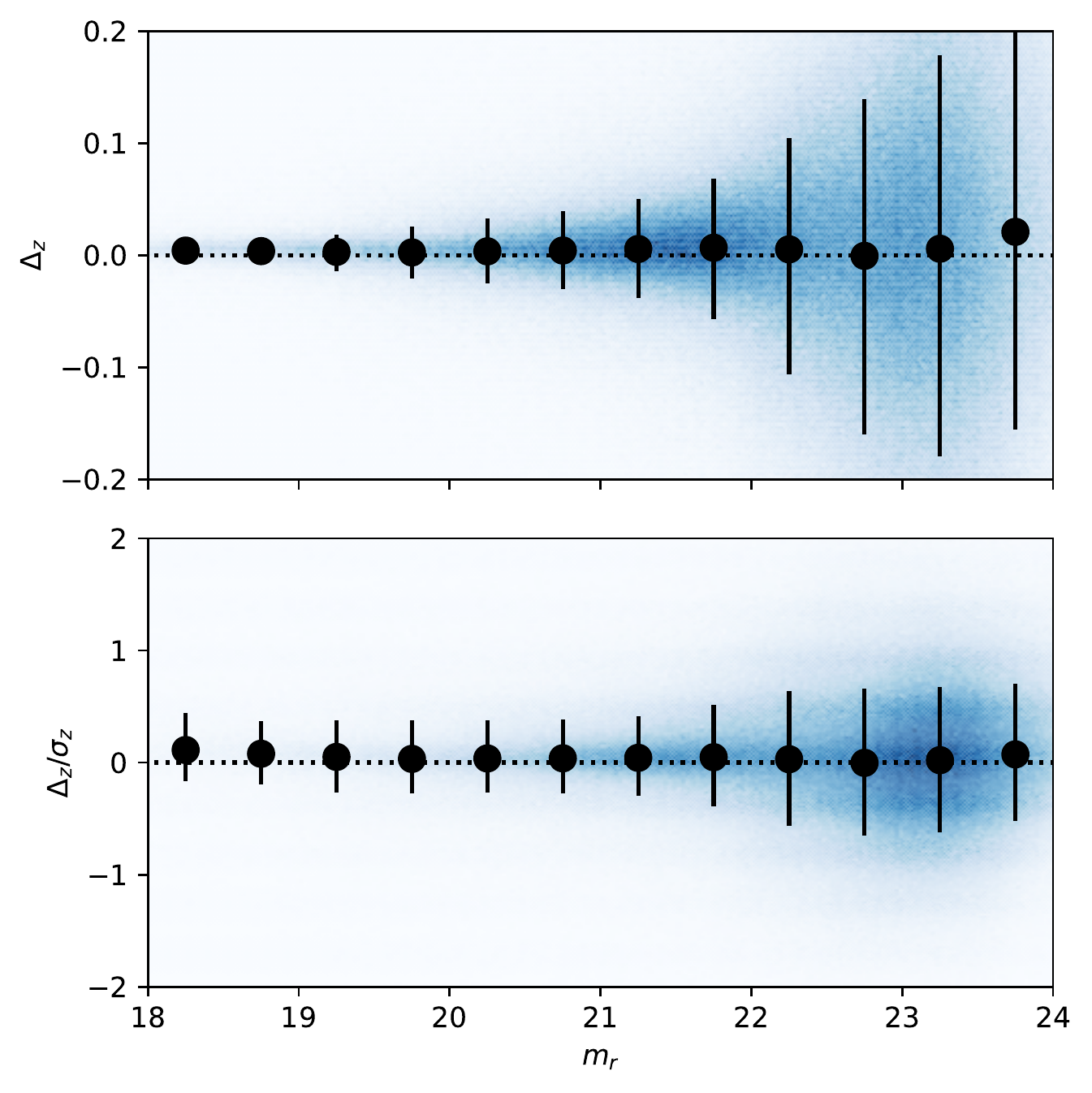}
 \caption{Difference in predicted photo-$z$ between LS DR8 North and South, as a function of apparent $r$-band magnitude, $m_{r}$. The upper panel shows the absolute photo-$z$ offset, while the lower panel shows the offset normalised by the combined photo-$z$ error (with errors added in quadrature).}
 \label{fig:north_south_offset_mag}
\end{figure}

\begin{figure}
\centering
 \includegraphics[width=0.98\columnwidth]{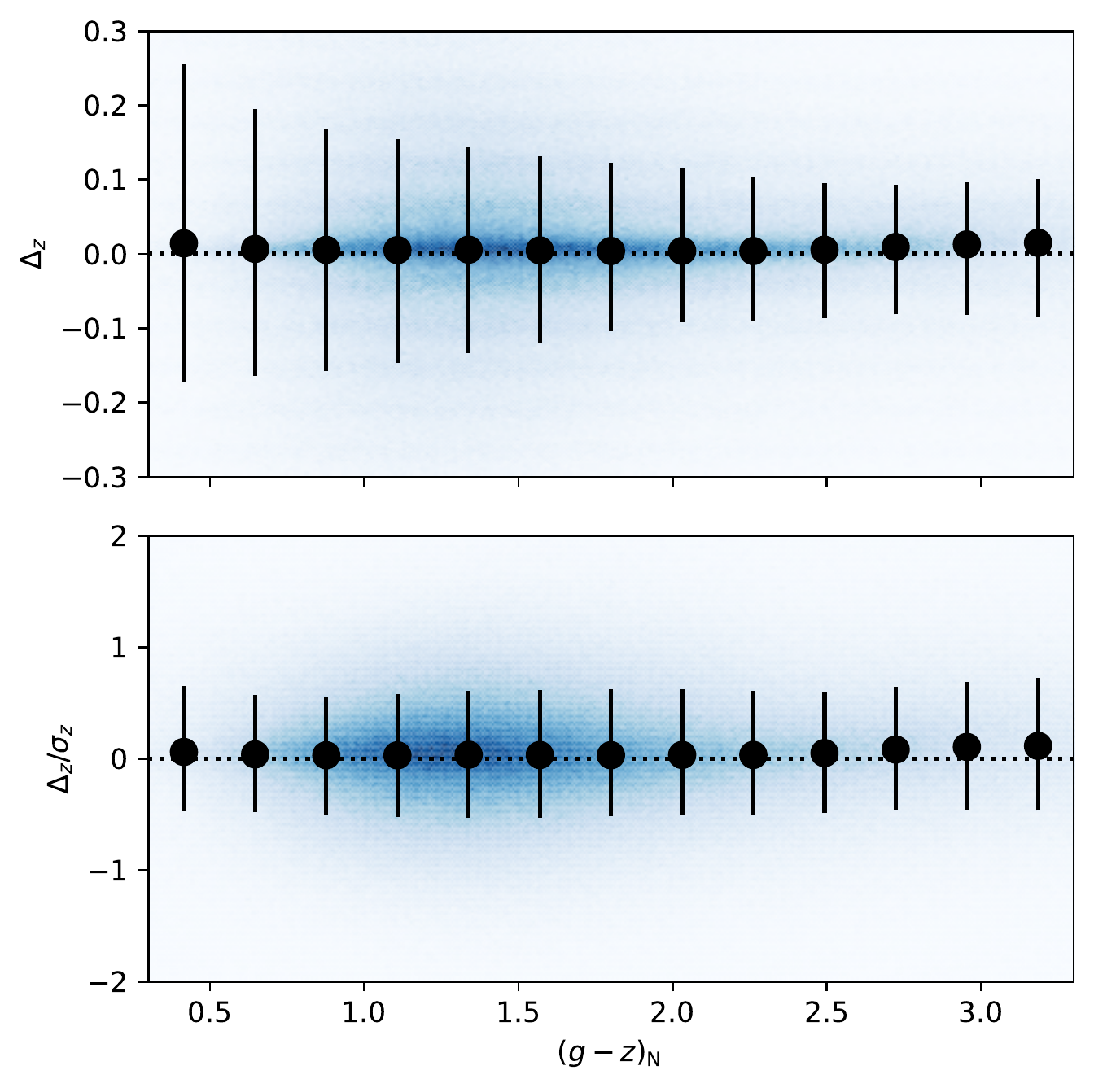}
 \caption{Difference in predicted photo-$z$ between LS DR8 North and South, as a function of optical colour, $g-r$ in the LS DR8 North photometry. The upper panel shows the absolute photo-$z$ offset, while the lower panel shows the offset normalised by the combined photo-$z$ error (with errors added in quadrature). Although the colour corrections are dependent on $g-z$, we find no evidence for systematic offsets in the resulting photo-$z$ predictions.}
 \label{fig:north_south_offset_col}
\end{figure}

\begin{figure}
\centering
 \includegraphics[width=0.98\columnwidth]{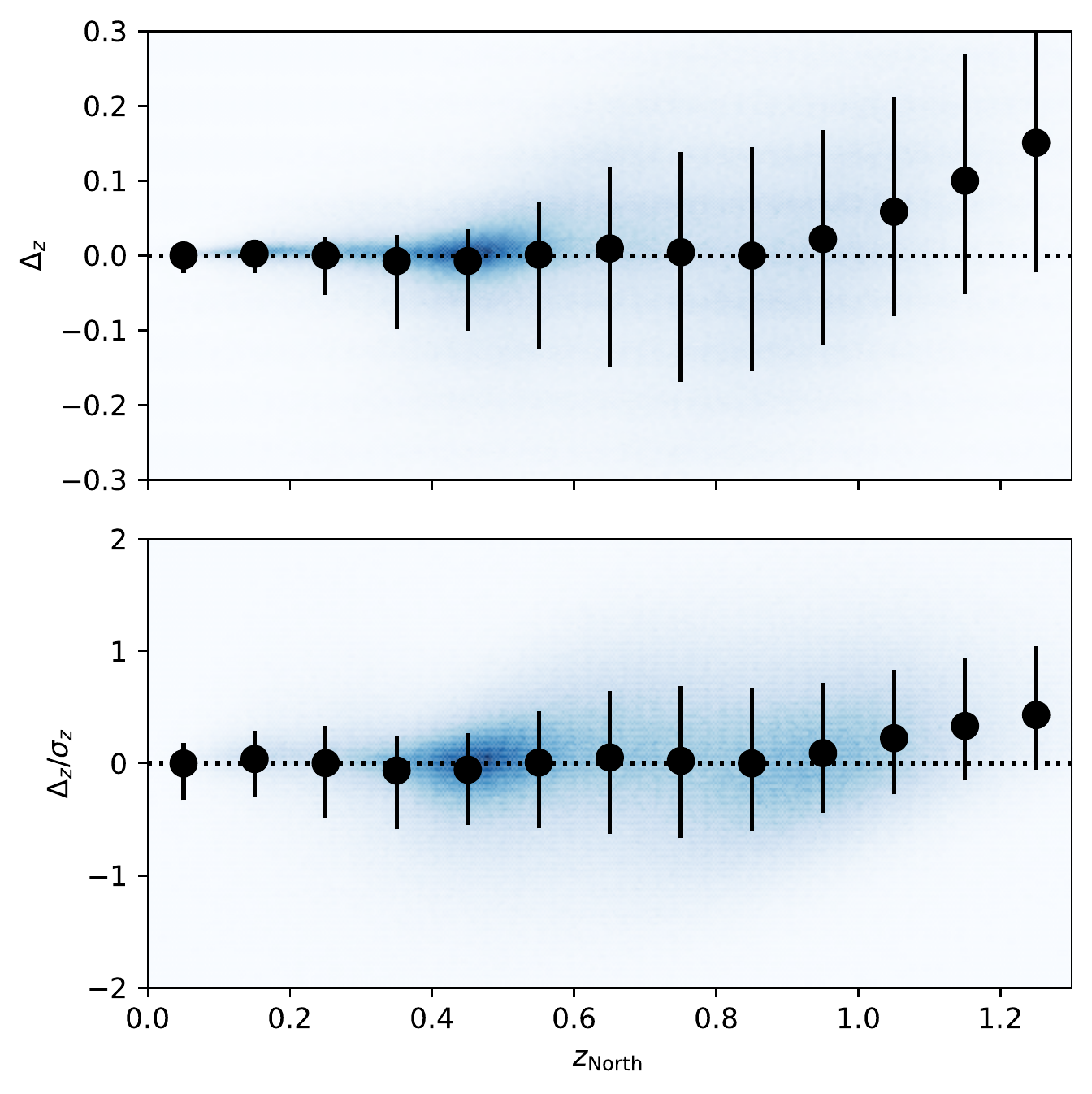}
 \caption{Difference in predicted photo-$z$ between LS DR8 North and South, as a function of photo-$z$, $z_{\text{North}}$. The upper panel shows the absolute photo-$z$ offset, while the lower panel shows the offset normalised by the combined photo-$z$ error (with errors added in quadrature).}
 \label{fig:north_south_offset_z}
\end{figure}

In Figs.~\ref{fig:north_south_offset_mag}, \ref{fig:north_south_offset_col} and \ref{fig:north_south_offset_z} we show the difference in photo-$z$ for the random sample as a function of apparent magnitude, colour and predicted redshift (LS DR8 North) respectively.
In all figures, the background colour distribution shows the density of sources with arbitrary scaling, with black circles and corresponding error bars showing the median and 16 to 84th percentiles for bins.
When examining both the absolute photo-$z$ offset, $\Delta_{z} = z_{\text{North}} - z_{\text{South}}$, and the error normalised offset, $\Delta_{z}/\sigma_{z}$ (where $\sigma_{z}^{2} = \sigma_{z, \text{North}}^{2} + \sigma_{z, \text{South}}^{2}$), we find no significant offset between the two predictions.
Across all magnitudes and colours, we find offsets that are $\ll 1\sigma$, demonstrating that the homogenisation step and any differences in softening parameter used to derive \emph{asinh} magnitudes have no significant impact on the resulting photo-$z$ predictions.
At $z > 1$, we do observe an increase in the difference between photo-$z$ predictions, but still remaining at below the $1\sigma$ level.
In this analysis we limit our comparison to $z < 1.3$ as our overlap sample does not provide sufficient statistics of high-$z$ sources beyond this redshift range.
However, in Section~\ref{sec:results_highz}, we find no evidence for systematic differences in the highest redshift sources present within the wider sample and are therefore confident that any systematic effects are negligible compared to the uncertainty on individual sources.


\bsp	
\label{lastpage}
\end{document}